\documentclass[showpacs,amsmath,amssymb,aps,twocolumn]{revtex4-1}
\usepackage{graphicx}
\usepackage{soul}
\usepackage[colorlinks=true,citecolor=blue,linkcolor=magenta]{hyperref}
\usepackage[usenames]{color}
\usepackage[cspex,bbgreekl]{mathbbol}
\usepackage{relsize}
\usepackage{printlen}

\newcommand{\si}{Supplementary Information}

\newcommand {\grsim} {\ {\raise-.5ex\hbox{$\buildrel>\over\sim$}}\ }
\newcommand {\lessim} {\ {\raise-.5ex\hbox{$\buildrel<\over\sim$}}\ } 
\newcommand {\ii} {i}

\def\be{\begin{equation}}
\def\ee{\end{equation}}
\def\bs#1{\boldsymbol{#1}}

\begin{document}

\bibliographystyle{naturemag}

\title{Measuring the Chern number of Hofstadter bands with ultracold bosonic atoms}

\author{M.~Aidelsburger$^{1,2}$, M.~Lohse$^{1,2}$, C.~Schweizer$^{1,2}$, M.~Atala$^{1,2}$, J.~T.~Barreiro$^{1,2,3}$, S.~Nascimb{\`e}ne$^{4}$, N.~R.~Cooper$^{5}$, I.~Bloch$^{1,2}$ \& N.~Goldman$^{4,6}$}

\affiliation{$^{1}$\,Fakult\"at f\"ur Physik, Ludwig-Maximilians-Universit\"at, Schellingstrasse 4, 80799 M\"unchen, Germany\\
$^{2}$\,Max-Planck-Institut f\"ur Quantenoptik, Hans-Kopfermann-Strasse 1, 85748 Garching, Germany\\
$^{3}$\,Present address: Department of Physics, University of California, San Diego, California 92093, USA\\
$^{4}$\,Coll\`{e}ge de France, 11 place Marcelin Berthelot $\&$ Laboratoire Kastler Brossel, CNRS, UPMC, ENS, 24 rue Lhomond, 75005 Paris, France\\
$^{5}$\,T.~C.~M.~Group, Cavendish Laboratory, J.J. Thomson Avenue, Cambridge CB3 0HE, United Kingdom \\
$^6$ CENOLI, Facult{\'e} des Sciences, UniversitŽ{\'e} Libre de Bruxelles (U.L.B.), B-1050 Brussels, Belgium}

\maketitle

\textbf{Sixty years ago, Karplus and Luttinger pointed out that quantum particles moving on a lattice could acquire an anomalous transverse velocity in response to a force, providing an explanation for the unusual Hall effect in ferromagnetic metals~\cite{Karplus:1954}. A striking manifestation of this transverse transport was then revealed in the quantum Hall effect~\cite{Klitzing:1986}, where the plateaus depicted by the Hall conductivity were attributed to a topological invariant characterizing Bloch bands: the Chern number~\cite{Thouless:1982}. Until now, topological transport associated with non-zero Chern numbers has only been revealed in electronic systems~\cite{Klitzing:1986,Dean:2013,Ponomarenko:2013}. Here we use studies of an atomic cloud's transverse deflection in response to an optical gradient to measure the Chern number of artificially generated Hofstadter bands~\cite{Hofstadter:1976}. These topological bands are very flat and thus constitute good candidates for the realization of fractional Chern insulators~\cite{Parameswaran:2013}. Combining these deflection measurements with the determination of the band populations, we obtain an experimental value for the Chern number of the lowest band $\nu_{\mathrm{exp}} =0.99(5)$. This result, which constitutes the first Chern-number measurement in a non-electronic system, is facilitated by an all-optical artificial gauge field scheme, generating uniform flux in optical superlattices.}

\begin{figure}[t!]
\includegraphics{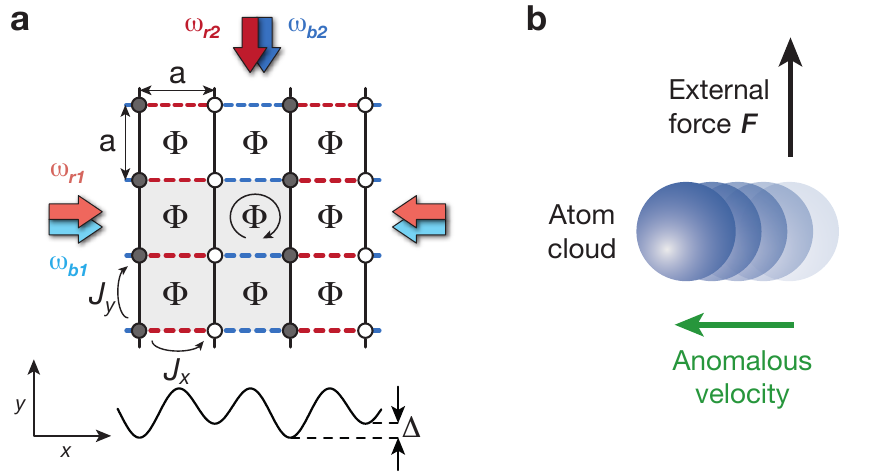}
\vspace{-0.cm} \caption{Schematics of the all-optical experimental setup used to generate a uniform artificial magnetic field and the Chern-number measurement. \textbf{a} The setup consists of a two-dimensional optical lattice with lattice constant $a=\lambda_s/2$ and tunnel couplings $J_x, J_y$ between neighboring sites. Bare tunneling is inhibited along $x$ by a staggered potential, creating an offset $\Delta$ between gray and white sites. Two additional pairs of laser beams (red and blue arrows), with wave vectors $|\mathbf{k}_{ij}|\simeq k_L=\pi/(2a)$  ($i=\{r,b\}$ and $j=\{1,2\}$) and resonant frequency difference $\omega_i=\omega_{i2}-\omega_{i1}=\pm\Delta/\hbar$, are used to restore  tunneling. Each pair consists of two beams, one running-wave (along $y$) and one retro-reflected beam (along $x$, arrows with lighter shading). Tunneling on red and blue links is controlled independently by the $i=r$ and $i=b$ beams, respectively, hence generating a rectified flux $\Phi=\pi/2$ per plaquette (aligned along $-\hat{\mathbf{e}}_z$). The magnetic unit cell (gray shaded area) is four times larger than the usual lattice unit cell.
\textbf{b} The Chern number is extracted from the transverse displacement of the atomic cloud, in response to an external force generated by an optical gradient. \label{Fig_Scheme}}
\end{figure}

\begin{figure*}[t]
\includegraphics{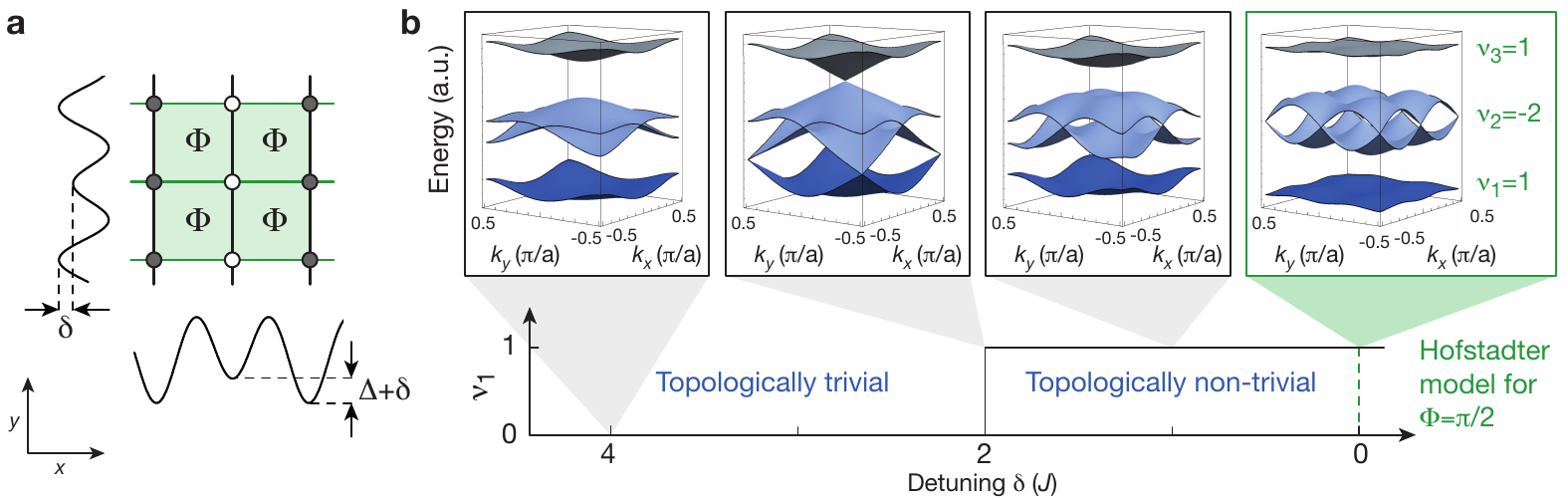}
\vspace{-0.cm} \caption{Energy spectra and topology of the Bloch bands as a function of a staggered detuning. \textbf{a} Schematic drawing of the Hofstadter-like optical lattice with an additional staggered detuning $\delta$ along $x$ and $y$. \textbf{b} Energy spectrum as a function of the staggered detuning. For a detuning larger than $2J$ the bands are topologically trivial with Chern numbers $\nu_{\mu}=0$. At the transition point, the band gaps close and the system enters a topologically non-trivial phase, where the Chern number of the lowest band $\nu_1=+1$. For vanishing detuning $\delta=0$, the system realizes the Harper-Hofstadter model with flux $\Phi=\pi/2$. Note that the vertical energy axis is rescaled for each spectrum.
\label{Fig:loading}}
\end{figure*}

One of the most challenging goals in the context of artificial gauge fields is to implement experimental probes revealing the non-trivial topology of energy bands. This would open the path towards the realization of novel topological states of matter in a wide class of physical settings, ranging from cold atoms to polariton gases~\cite{Goldman:2013,Bermudez:2011,Rechtsman:2013,Carusotto:2013}. Considering cold atoms in optical lattices, it has been shown theoretically that transport measurements could allow for a determination of the Chern number characterizing topological Bloch bands~\cite{Price:2012,Dauphin:2013,Goldman:2013}. Although earlier experiments used local cyclotron orbits to detect the artificial gauge field structure at the single-plaquette level~\cite{Aidelsburger2011a,Aidelsburger:2013}, observing the entire cloud dynamics or determining the Chern number of the underlying bands has remained out of reach. Here we load ultracold bosonic atoms into Hofstadter bands, and we report on the direct detection of the cloud's spatial motion as a response to an applied force. Using a novel band-mapping technique, we track the populations of the Hofstadter bands over time. We observe that the flatness of the bands leads to homogeneous populations within individual bands, through interactions and finite temperatures. For short times, heating and non-adiabatic interaction effects are limited, and the detected transverse Hall drift is in agreement with exact diagonalization studies~\cite{Price:2012,Dauphin:2013}. Combined with independent band-population measurements, we extract the Chern number of the lowest Hofstadter band from our experimental data. Our work represents the first determination of a topological invariant characterizing two-dimensional Bloch bands using ultracold atoms, and complements other studies based on atom interferometric~\cite{Atala:2013,Abanin:2013} and collective-mode~\cite{LeBlanc:2012} measurements.

Our optical-lattice setup realizes the celebrated Harper-Hofstadter Hamiltonian~\cite{Hofstadter:1976}, which describes the motion of particles on a square lattice penetrated by a uniform magnetic flux $\Phi$ per unit cell [see also the original works by Harper~\cite{Harper:1955} and Azbel~\cite{Azbel:1964}]. An atom that hops around a closed loop of the optical lattice picks up a geometric phase, which mimics the Aharonov-Bohm phase of a charged particle in a magnetic field. The artificial flux is thus produced by engineering complex tunneling matrix elements with space-dependent (Peierls) phases~\cite{Hofstadter:1976}, through the laser-assisted-tunneling method introduced by Jaksch and Zoller~\cite{Jaksch:2003} and subsequent works~\cite{Gerbier:2010,Mueller:2004,Kolovsky:2011,Bermudez:2011, Baur:2014}. The present scheme controls tunneling along the $x$ direction and uses two main ingredients: a staggered potential with energy offset $\Delta$ between neighboring sites, inhibiting bare tunneling along $x$, together with a resonant modulation of frequency $\omega = \Delta / \hbar$ restoring the hopping in a controlled way. Using a simple time-dependent optical potential, which simultaneously triggers the hopping on all the links, naturally produces a staggered flux distribution~\cite{Aidelsburger2011a}. In order to rectify the flux, we implement a new all-optical driving scheme that uses two pairs of laser beams to control the laser-induced tunneling on successive links independently, hence producing a uniform flux $\Phi=\pi/2$ per unit cell (Fig.~\ref{Fig_Scheme}a). The lowest band of the corresponding energy spectrum has a Chern number $\nu_1=+1$ and is thus topologically equivalent to the lowest Landau level. Moreover, this band exhibits a large flatness ratio of $E_{\mathrm{gap}}/E_{\mathrm{bw}}\simeq7$, where $E_{\mathrm{gap}}$ is the energy gap to the second band and $E_{\mathrm{bw}}$ the bandwidth. In contrast to previous experiments generating uniform flux in optical lattices~\cite{Aidelsburger:2013,Miyake:2013}, the present scheme does not rely on magnetic field gradients, and therefore offers a higher degree of experimental control.

The experimental setup consists of an ultracold gas of $^{87}$Rb atoms that is loaded into a two-dimensional lattice created by two orthogonal standing waves with wavelength $\lambda_s=767\,\mathrm{nm}$. The atoms are confined in the perpendicular direction by a weak harmonic potential using an optical dipole trap. An additional standing wave with twice the wavelength $\lambda_L=2\lambda_s$ is superimposed along $x$ to create the staggered potential (Fig.~\ref{Fig_Scheme}a), with an energy offset $\Delta$ much larger than the bare tunneling $J_x$. The modulation restoring resonant tunneling is created by two additional pairs of far-detuned laser beams,  each pair generating a moving potential of the form $V_i(x,y,t)=\kappa\ \mathrm{cos}(k_L x + \varphi_i) \mathrm{cos}(-k_L y \pm \omega t)$, where $\kappa$ is the driving amplitude, $k_L=2\pi/\lambda_L$, and $\omega = \Delta / \hbar$. The relative phases $\varphi_i$ are adjusted so as to control successive links independently (Fig.~\ref{Fig_Scheme}a). In the high-frequency limit $\hbar \omega \gg J_x, J_y$, this system can be described by an effective time-independent Hamiltonian~\cite{Sorensen:2005,Goldman:2014,Bukov:2014,Lignier:2007,Struck:2011}, whose dominant contributions reproduce the Harper-Hofstadter Hamiltonian (see \si)

\def\pM{\mathrel{\raise -1.6pt \hbox{\tiny(}\! 
                 \raise 1.8pt \hbox{+}
                 \settowidth {\dimen03} {+}
                 \hskip-\dimen03
                 \raise -2.5pt \hbox {$-$}
                 \!\raise -1.6pt \hbox{\tiny)}}}

\begin{eqnarray}
&&\hat H=-J\sum_{m,n}\left(\mathrm{e}^{ \ii n\Phi}\hat a^{\dagger}_{m+1,n}\hat a^{\phantom{\dagger}}_{m,n}
                     +\hat a^{\dagger}_{m,n+1}\hat a^{\phantom{\dagger}}_{m,n}+\mathrm{h.c.}\right), \nonumber \\  
                    && \Phi= \pi/2,
                      \label{eq:effectiveHam}
\end{eqnarray}

\noindent where the Landau gauge was chosen to describe the system~\cite{Hofstadter:1976}. Here $\hat a^{\phantom{\dagger}}_{m,n} (\hat a^{\dagger}_{m,n})$ annihilates (creates) a particle on site $(m,n)$, where the position in the lattice is defined as $\mathbf{R}=ma\hat{\mathbf{e}}_x + na\hat{\mathbf{e}}_y$, with $m,n$ integers and $\hat{\mathbf{e}}_{x,y}$ the unit vectors. In the limit $\Delta\gg \kappa$, the effective coupling strengths are given by $J^{\mathrm{eff}}_x\simeq J_x \kappa/(\sqrt{2}\Delta)$ and $J^{\mathrm{eff}}_y\simeq  J_y$; the experimental parameters were chosen such that $J^{\mathrm{eff}}_x\approx J^{\mathrm{eff}}_y \equiv J$. Higher order corrections to the effective Hamiltonian lead to a local renormalization of the hopping along $y$, which for our experimental parameters $\kappa/(\hbar\omega)$ can lead to modifications of the tunneling up to $0.3 J_y$ (see \si). In the presence of the effective flux $\Phi=\pi/2$, the magnetic unit cell is four times larger than the standard unit cell (Fig.~\ref{Fig_Scheme}a). Consequently the first magnetic Brillouin zone (FBZ) is reduced, and the energy bands split into four subbands~\cite{Hofstadter:1976,Thouless:1982}. Since the two middle bands touch at the Dirac points (Fig.~\ref{Fig:loading}b), the energy spectrum is partitioned into three isolated bands, labeled as $E_{\mu}$, with Chern numbers $\nu_{\mu}$. We stress that the central ``super-band'' contains twice the number of states as compared to the other two bands.

In order to load the atoms into the lowest Hofstadter band, we developed an experimental sequence using an auxiliary superlattice potential (see \si), which introduces a staggered detuning $\delta$ along both directions: along $x$, the offset between neighboring sites is increased away from the resonance condition to $\Delta+\delta$, while it is simply given by $\delta$ along $y$ (Fig.~\ref{Fig:loading}a). Importantly, the unit cell of the square lattice with staggered potentials along both directions is equivalent to the magnetic unit cell of the Harper-Hofstadter model, and thus, the number of energy bands is preserved during the loading sequence. For $\delta > 2J$ the topology of the bands is trivial and all Chern numbers are zero. When crossing the topological phase transition at $\delta=2J$, the spectral gaps close at a single point in the FBZ and the system enters the topologically non-trivial regime, where the lowest band $E_1$ has a Chern number $\nu_1=+1$. Finally, for $\delta=0$ we realize the Harper-Hofstadter model with flux $\Phi=\pi/2$ (Fig.~\ref{Fig:loading}b).

\begin{figure*}[t!]
\includegraphics{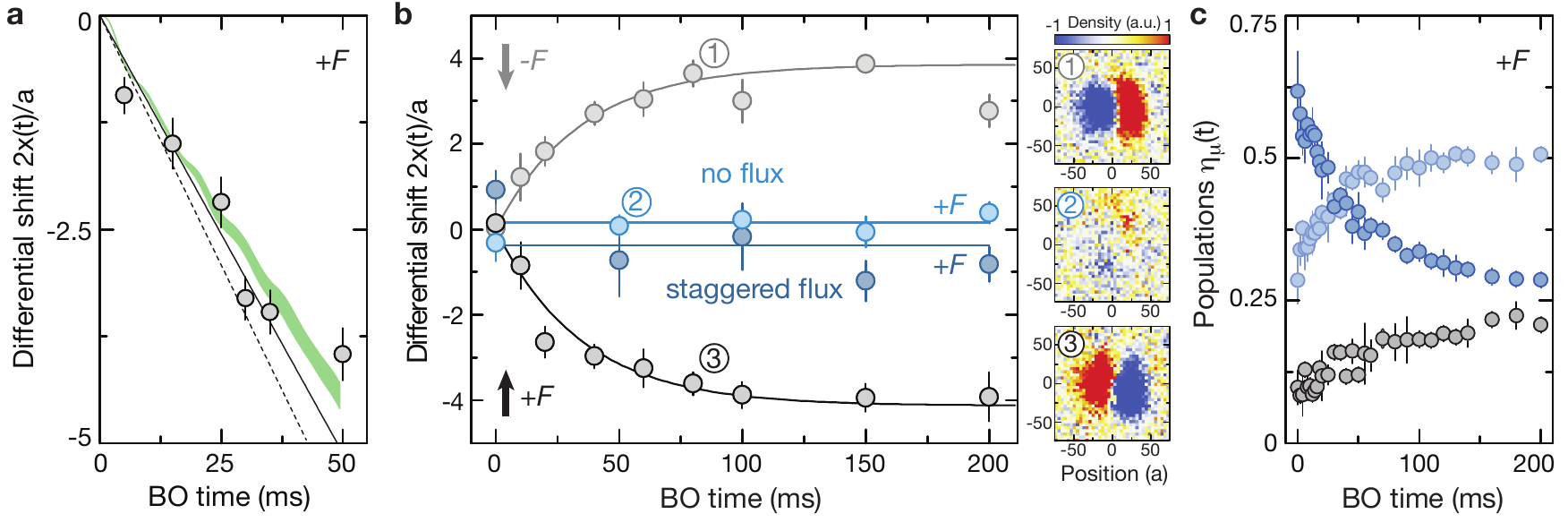}
\vspace{-0.cm} \caption{Differential center-of-mass displacements and band populations $\eta_{\mu} (t)$ versus Bloch oscillation (BO) time. The differential shifts were measured after loading the atoms in the Hofstadter bands ($\Phi=\pi/2$, $J/h=75(3)\,$Hz) and applying an external force $\mathbf{F}=\pm F \hat{\mathbf{e}}_y$ ($Fa/h=38.4(8) \mathrm{Hz}$). The value of the differential shift was evaluated after averaging ten images per sign of the flux $\pm \Phi$ and subsequently subtracting the COM-positions of the atomic cloud. Henceforth we define this as one dataset. Each data point is an average of five datasets and the error bars depict the error of the mean. 
\textbf{a} The black solid line is a linear fit to the data for $t\leq 35\,$ms. Taking into account the measured initial band populations $\eta^0_{\mu}=\{0.55(6),0.31(3),0.13(3)\}$ and using equation~(\ref{eq:driftfraction}) yields the value $\nu_{\mathrm{exp}}=0.9(2)$. The dashed line depicts the ideal evolution for $\nu_1=1$ and the same initial band populations. The green shaded area illustrates the corresponding numerical simulations. These numerics are based on the Hamiltonian in equation~(\ref{eq:effectiveHam}) supplemented by small higher-order corrections (see \si). 
\textbf{b} The black (gray) data points show the evolution for longer times and opposite directions of the applied force $\mathbf{F}=\pm F \hat{\mathbf{e}}_y$. The small images on the right show typical experimental data obtained after subtracting the averaged images of one dataset for $\pm \Phi$. The blue data points were taken in lattice configurations with trivial topology, $\Phi=0$ (light blue) and $\Phi=(-1)^m\pi/2$ (dark blue); the blue solid lines are guides to the eye. The solid black and gray lines show a fit to our data using equation~(\ref{eq:driftfraction_int}) combined with the filling factor $\gamma(t)$, which was evaluated using the measured band populations $\eta_{\mu} (t)$ shown in (c) and fitted with an exponential. This provides an experimental Chern number $\nu_{\mathrm{exp}}=1.05(12)$ (black) and $\nu_{\mathrm{exp}}=0.98(12)$ (gray).
\textbf{c} Evolution of the band populations $\eta_{\mu} (t)=N^{(\mu)}/N_{\mathrm{tot}}$, defined as the fraction of particles in band $\mu$. Each data point is an average of two individual measurements and the error bars denote the standard deviation of the data points. The color code illustrates the connection to the Hofstadter bands shown in Fig.~\ref{Fig:loading}b.  \label{Fig:hallResponse}}
\end{figure*}

Recently, several methods were proposed to probe the topological nature of energy bands with cold atoms, exploiting Bloch oscillations and other transport measurements~\cite{Price:2012,Dauphin:2013,Goldman:2014}. In the presence of a constant force $\mathbf{F}=F \hat{\mathbf{e}}_y$, atoms on a lattice undergo Bloch oscillations along the direction of the force; this longitudinal motion is captured by the band velocity ${\mathbf v}_{\mu}^{\rm{band}} = \partial_{{\bf k}} E_{\mu}/\hbar$. Moreover, when the energy bands have nonzero Berry curvature, the cloud also experiences a net perpendicular (Hall) drift (Fig.~\ref{Fig_Scheme}b); this transverse motion is described by an additional contribution to the velocity, denoted $v^x_{\mu}\,$\cite{Xiao:2010}. For a particle in a state $\left|u_{\mu}(\mathbf{k}) \right>$ of the $\mu$-th band, this ``anomalous" contribution to the velocity reads

\begin{eqnarray}
&&v^x_{\mu}(\mathbf{k}) = -\frac{F}{\hbar} \Omega_{\mu}(\mathbf{k}), \\
&&\Omega_{\mu} = i \left ( \left< \partial_{k_{x}} u_{\mu} | \partial_{k_{y}} u_{\mu} \right> - \left< \partial_{k_{y}} u_{\mu} | \partial_{k_{x}} u_{\mu} \right> \right ), \nonumber
\end{eqnarray}

\noindent where $\Omega_{\mu} (\mathbf{k})$ is the Berry curvature of the band~\cite{Xiao:2010}. The effects associated with the anomalous velocity $v^x_{\mu}$ can be isolated by uniformly populating the bands, which averages out any contribution from the band velocity, $\int \partial E_{\mu}/\partial k_{x,y} \mathrm{d}^2\mathrm{k}=0$. This could be directly realized with fermionic atoms by setting the Fermi energy within a spectral gap~\cite{Dauphin:2013}, in analogy with the integer quantum Hall effect. Here we consider an incoherent distribution of bosonic atoms, where the population within each band is homogeneous in $k$-space, an assumption which has been validated independently (see \si). In the absence of inter-band transitions, the contribution of the $\mu$-th band to the center-of-mass (COM) motion perpendicular to the force can be written in terms of the Chern number of the $\mu$-th band $\nu_{\mu}=\int_{\mathrm{FBZ}} \Omega_{\mu} \mathrm{d}^2\mathrm{k}/(2\pi)$,

\begin{equation}
x_{\mu}(t)=-\frac{4a^2 F}{h}\,\nu_{\mu} \, t = - 4a \,\nu_{\mu} \, \frac{t}{\tau_B},\
\label{eq:drift}
\end{equation}

\noindent where the factor $4a^2$ corresponds to the extended unit cell (see Fig.~\ref{Fig_Scheme}a) and $\tau_B = h/(Fa)$ is the characteristic time scale for Bloch oscillations. In our experiments, we applied an optical dipole force along $y$ (\si) and measured the COM-evolution of the atomic cloud in-situ for opposite directions of the flux $\Phi$, which were then subtracted to obtain the differential shift $x(t,\Phi)-x(t,-\Phi)=2 x(t)$. For short evolution times, an almost linear differential displacement is observed (Fig.~\ref{Fig:hallResponse}a), while for longer times it saturates due to band repopulation (Fig.~\ref{Fig:hallResponse}b,c). We note that the deflection is symmetric with respect to the direction of the applied force (black and gray data points in Fig.~\ref{Fig:hallResponse}b), as expected from theory. Additionally, we measured the COM-motion for $\Phi=0$ (light blue data points in Fig.~\ref{Fig:hallResponse}b) and for a staggered-flux distribution (dark blue data points in Fig.~\ref{Fig:hallResponse}b). Both measurements do not show any significant displacement, which is consistent with a Chern number of zero (see \si).

The band-mapping sequence, which is basically the reversed loading sequence as illustrated in Fig.~\ref{Fig:loading}b, allows us to measure the band populations of the different Hofstadter bands during the dynamics (Fig.~\ref{Fig:hallResponse}c). For large detuning $\delta$, tunneling is inhibited along both directions and the populations of the Hofstadter bands map onto the ones of the two-dimensional superlattice, where standard detection techniques can be used to evaluate the band populations $\eta_{\mu}$ (see \si)~\cite{Nascimbene:2012}. The contribution of all atoms in different bands to the mean COM-displacement can be written as
\begin{equation}
x(t)= -4a \gamma_0 \, \nu_1 \, \frac{t}{\tau_B}, \ \mathrm{with} \  \gamma_0=\eta^0_1-\eta^0_2+\eta^0_3 \ ,
\label{eq:driftfraction}
\end{equation}

\begin{figure}[t!]
\includegraphics{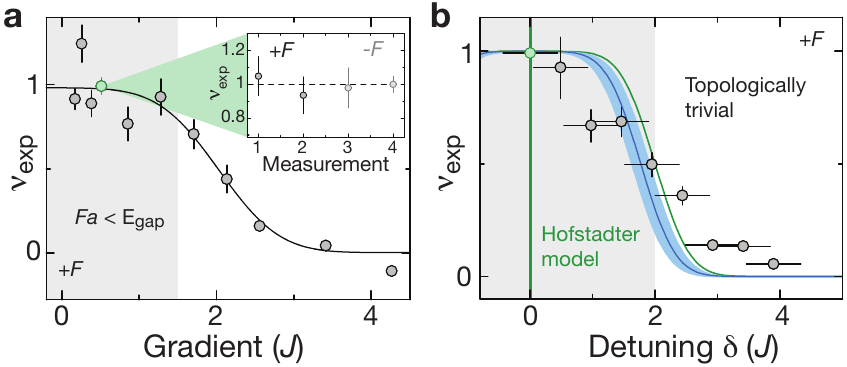}
\vspace{-0.cm} \caption{Measured Chern number $\nu_{\mathrm{exp}}$ as a function of gradient strength $Fa$ and staggered detuning $\delta$.
\textbf{a} For each value of the gradient, we measured the full time evolution of the differential shift and filling factor $\gamma (t)$ (similar to the evolution shown in Fig.~\ref{Fig:hallResponse}b,c) to extract the value of the Chern number. The solid line is a guide to the eye to highlight the saturation to $\nu_{\mathrm{exp}}\approx 1$ for small gradients. The green data point depicts the experimentally determined Chern number $\nu_{\mathrm{exp}}=0.99(5)$ for $\pm Fa/h=38.4(8) \mathrm{Hz}$, for which a larger number of measurements were taken as displayed in the inset.
\textbf{b} The differential shift versus detuning was measured after four different times $t=(20, 50, 100, 150)\,$ms for $Fa/h=38.4(8) \mathrm{Hz}$ to fit the Chern number based on the measured filling factor $\gamma (t)$. The topological phase transition, which is expected at $\delta=2J$ for the model (Fig.~\ref{Fig:loading}b), is smoothened due to experimental uncertainties (green solid curve). The blue shaded area illustrates the range of transition points $\delta=1.77(14) J$ given by the second-order correction, for our experimental parameters $\kappa/(\hbar\omega)=0.58(2)$ (see \si). The green and blue curves are calculated taking into account the experimental uncertainty in the detuning ($0.4 J$, horizontal error bars), which is determined experimentally by typical drifts and fluctuations of the resonance frequency. All data points include an average of five datasets for each time-step for the differential shift and two measurements for $\gamma(t)$. As in (a) the green data point depicts the experimentally determined Chern number $\nu_{\mathrm{exp}}= 0.99(5)$ for $\delta=0$. All vertical error bars display the uncertainty resulting from the fit error of $\gamma (t)$. \label{Fig:det_grad}}
\end{figure}

\noindent where the filling factor $\gamma_0$ is given by the initial band populations $\eta_{\mu}^0$. This result is obtained by invoking the particle-hole symmetry inherent to the Harper-Hofstadter model (i.e. $\nu_1 = \nu_3$), together with the fact that the sum of Chern numbers necessarily vanishes, $\sum_{\mu} \nu_{\mu} =0$; we have also taken into account that the middle band contains twice the number of states as compared to the other two bands; its contribution to $x(t)$ is thus proportional to $\nu_2 (\eta_2/2) = - \nu_1 \eta_2$ (see \si). As a result, the COM displacement in equation~(\ref{eq:driftfraction}) is only determined by the initial band populations $\eta_{\mu}^0$ and the Chern number of the lowest band $\nu_1$, under the assumption that the band populations are constant over time, which is reasonable for short times. Comparing the short-time trajectories of the atomic cloud $x(t)$ with the equation of motion~(\ref{eq:driftfraction}), together with the measured initial filling factor $\gamma_0$, provides a first reasonable experimental value for the Chern number of the lowest band $\nu_{\mathrm{exp}}=0.9(2)$. In particular, for short times, we find good agreement between the theoretical predictions based on the effective Hamiltonian and the experimental data (Fig.~\ref{Fig:hallResponse}a).

We now present a more precise Chern-number measurement based on a long-time analysis, which takes into account the repopulation of atoms between the three Hofstadter bands (see Fig.~\ref{Fig:hallResponse}c). One possible reason for this are Landau-Zener transitions, which are neglected in equation~(\ref{eq:driftfraction}) but well captured by the numerical simulations (green shaded area in Fig.~\ref{Fig:hallResponse}a). However, we observe similar repopulation timescales in the absence of the force, most likely due to heating of the atoms caused by the periodic driving. In order to capture the band repopulation effects, we benefit from the measured filling factor $\gamma(t)=\eta_1(t)-\eta_2(t)+\eta_3(t)$ and model the dynamics according to the modified equations of motion
\begin{equation}
x(t)=-4 a \, \nu_1 \, \int_0^t \gamma(t')  \mathrm{dt}'/\tau_B  \ .
\label{eq:driftfraction_int}
\end{equation}

\noindent By fitting this equation to the experimental data $x(t)$, with the Chern number being the only fit parameter, we obtain an experimental value for the Chern number of the lowest band
\begin{equation}
\nu_{\mathrm{exp}}= 0.99(5) \ .
\label{eq:CNexp}
\end{equation}

\noindent Here we averaged over four independent Chern-number measurements, two for each direction of the gradient to avoid systematic errors (see inset of Fig.~\ref{Fig:det_grad}a). The stated uncertainty is the standard deviation obtained from these measurements. This shows that including our time-resolved band-mapping data into our modeling of the transverse Hall drift leads to a very good understanding of the full time dynamics, and allows us to extract the value of the Chern number with high accuracy. The value of the applied force was chosen to be large enough to accurately detect the displacement, but weak enough to limit non-linear effects and Landau-Zener induced inter-band transitions.

The dependence of our Chern-number measurement with respect to the force was studied in more detail, as shown in Fig.~\ref{Fig:det_grad}a. For gradient strengths smaller than the band gap, $Fa<E_{\mathrm{gap}}\approx 1.5J$, the experimental value for the Chern number saturates to $\nu_{\mathrm{exp}}\approx 1$, indicating that it can be determined reliably for small forces. For very strong forces $Fa > E_{\mathrm{gap}}$, our model breaks down and the experimental value $\nu_{\mathrm{exp}}$ decreases to zero.

Finally we characterized the topological phase transition, which is expected for a staggered detuning of $\delta=2J$ (see Fig.~\ref{Fig:loading}). For this analysis, we set the gradient amplitude to the value $Fa=38.4(8) \mathrm{Hz} \times h=0.51(1)J$, which is well below the band gap for $\delta=0$. In agreement with theory, we observe that the experimental value for the Chern number decreases to zero across the phase transition (Fig.~\ref{Fig:det_grad}b). We note that Landau-Zener transitions to higher bands become more important when approaching the transition (gap-closing point); however, this should not affect the measurement since the band-repopulation is taken into account, according to equation~(\ref{eq:driftfraction_int}). The smoothened transition is most likely due to the experimental uncertainties in the resonance condition (green solid line in Fig.~\ref{Fig:det_grad}b). In addition, second-order corrections to the effective Hamiltonian shift the transition point to a mean value of $\delta \approx 1.8J$ for our experimental parameters $\kappa/(\hbar\omega)=0.58(2)$ (solid blue line and shaded region in Fig.~\ref{Fig:det_grad}b, see \si).

In conclusion, we have successfully implemented a method to measure the Chern number in a cold-atom setup, which can be generalized to a wide range of non-electronic systems, including ion traps~\cite{Bermudez:2011}, photonic crystals~\cite{Rechtsman:2013} and polaritons~\cite{Carusotto:2013}. While our  measurement accommodates dynamical transitions to higher bands, which we attribute to the lattice modulation used to engineer the topological band structure, our results highlight the necessity to further deepen the understanding of heating processes in periodically-driven quantum systems. Minimizing heating effects and clarifying the role of interactions in these modulated systems will be crucial in view of reaching topological strongly-correlated states in Chern bands, such as fractional Chern insulators~\cite{Parameswaran:2013}.\\
\newline
Recently we have become aware of related measurements showing signatures of the Berry curvature in periodically modulated honeycomb optical lattices~\cite{Jotzu:2014}.\\
\newline
We acknowledge fruitful discussions with J.~Dalibard and also with A.~Dauphin, P.~Gaspard, F.~Gerbier, F.~Grusdt, I.~Carusotto, T.~Ozawa and H.~Price. This work was supported by NIM, the EU (UQUAM, SIQS) and the EPSRC Grant No. EP/K030094/1. M.~Ai. was additionally supported by the Deutsche Telekom Stiftung, M.~L. by ExQM and N.~G. by the Universit\'{e} Libre de Bruxelles and the FRS-FNRS (Belgium).

\newpage

\section*{\si}
\twocolumngrid

\renewcommand{\thefigure}{S\arabic{figure}}
 \setcounter{figure}{0}
\renewcommand{\theequation}{S.\arabic{equation}}
 \setcounter{equation}{0}
 \renewcommand{\thesection}{S.\Roman{section}}
\setcounter{section}{0}

\section{Flux rectification in a staggered optical potential}
\label{sec:rectification}

\begin{figure}[thb]
\begin{center}
\includegraphics[scale=1]{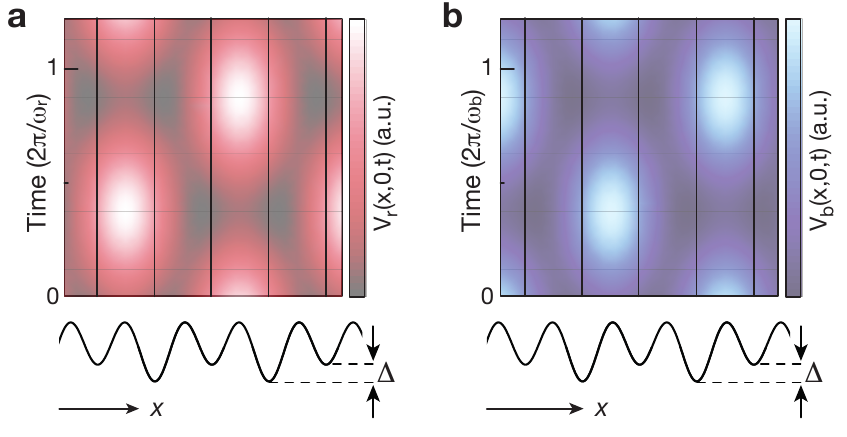}
\caption{Total time-dependent potential $V_r(x,0,t)$ \textbf{(a)} and $V_b(x,0,t)$ \textbf{(b)} as a function of time for $y=0$. For each of the two modulations (red and blue) the relative modulation amplitude between neighboring sites vanishes on every other bond. The modulation depicted in (a) therefore induces tunneling only on bonds with a positive energy offset $+\Delta$ between neighboring sites and the one depicted in (b) on bonds with negative energy offset $-\Delta$.}\label{fig:S1}
\end{center}
\end{figure}

In this section we describe the laser configuration used to rectify the artificial magnetic field in a two-dimensional superlattice potential as illustrated in Fig.~1a of the main text. The staggered energy offset $\Delta$ introduced by the staggered superlattice potential inhibits tunneling along $x$ for $\Delta\gg J_x$, with $J_x$ being the bare tunnel coupling. In our previous work~\cite{aidelsburger2011experimental} we used one pair of far-detuned running-wave beams to restore the tunneling along this direction, which results in a staggered effective magnetic field whose sign is alternating along $x$. In order to rectify this artificially generated magnetic field we now employ two pairs of beams (red and blue arrows in Fig.~1a of the main text), where each of them addresses only every other bond (Fig.~\ref{fig:S1}), such that the sign of the Peierls phases on the two different kinds of bonds (red and blue dashed lines in Fig.~1a of the main text) can be controlled individually. Each pair consists of two beams, along the $x$- and $y$-direction, where the ones along $x$ are retro-reflected creating a standing-wave that interferes with the running-wave along $y$. The corresponding local time-dependent optical potential is given by
\begin{eqnarray}
V_i(x,y,t) &=& 4 E_{i1}^2 \textrm{cos}^2(k_L x +\varphi_i) + E_{i2}^2 \nonumber \\
&+& 4 E_{i1} E_{i2} \textrm{cos}(k_L x + \varphi_i) \times \nonumber \\
&&\textrm{cos}(- k_L y + \omega_i t + \phi_i) \  ,
\label{eq:Vxy}
\end{eqnarray}

\noindent where $\varphi_i$ is the phase relative to the underlying lattice and $\phi_i$ is the phase of the modulation, $i=\{r,b\}$. The local potential consists of two parts, a static standing-wave term with constant offset and a time-dependent interference term. In Fig.~\ref{fig:S1} we show the total time-dependent potential $V_i$ for $y=0$, $\varphi_r=-\pi/4$ and $\varphi_b=\pi/4$. It illustrates that for an appropriate choice of $\varphi_i$ the relative modulation between neighboring sites vanishes on every other bond. Furthermore if $\varphi_b=\varphi_r+\pi/2$ the two pairs of beams address two different kinds of bonds with positive and negative sign of the energy offset $\pm\Delta$ between neighboring sites. This has the additional advantage that the two standing wave-terms in $V_r(x,y,t)$ and $V_b(x,y,t)$ cancel each other.

In our experimental setup all four beams are realized using a single laser. The beam is split into two parts (beam 1 $\&$ 2). Subsequently each of them is sent through a fiber-coupled intensity modulator, which is used to create two sidebands (red and blue) with frequencies $\omega_{rj,bj}$, $j=\{1,2\}$. The carrier frequency is fully suppressed. The frequency differences between each pair of sidebands are given by $\omega_{r1}-\omega_{b1}=2 \pi \times185\,$MHz and $\omega_{r2}-\omega_{b2}=2 \pi \times 185\,$MHz$+ \Delta / \hbar$. Due to the large frequency difference between the sidebands $\omega_{rj,bj}$ we can neglect the corresponding interference terms, such that the only relevant time-dependent terms for the modulation are given by the interference between $\omega_{i1}$ and $\omega_{i2}$. Since the sidebands are generated symmetrically around the carrier frequency, they also have the same amplitudes $E_{rj}=E_{bj}\equiv E_j$. This beam configuration thus leads to a local time-dependent optical potential of the form

\begin{eqnarray}
V_{m,n}(t) &=& \kappa\ \textrm{cos} (m \pi /2  -\pi/4 ) \times \label{eq:mod} \\ \nonumber
&&\textrm{cos}(- n \pi / 2 + \omega_r t +\phi_{r} ) \\ \nonumber 
&+& \kappa\ \textrm{cos}(m \pi/2  +\pi/4 ) \times \\ \nonumber
&&\textrm{cos}(-n \pi/2 + \omega_b t+ \phi_{b} ) \ ,
\end{eqnarray}

\noindent with $\kappa=4 E_{1} E_{2}$ and $\omega_i=\omega_{i2}-\omega_{i1}$. The position in the lattice is defined as $\mathbf{R}=ma\hat{\mathbf{e}}_x + na\hat{\mathbf{e}}_y$, with $m,n$ integers and $\hat{\mathbf{e}}_{x,y}$ the unit vectors. The relative phase between the two modulations $\phi_b-\phi_r$ can be controlled in the experiment but its value neither influences the value of the effective flux realized by the modulation nor the strength of the effective couplings that appear in the effective Hamiltonian. Therefore, without loss of generality we choose this phase equal to $\phi_b-\phi_r=\pi/2$. The overall phase of the modulation relative to the underlying lattice however is random because the phase of the running-wave along $y$ is not stabilized with respect to the lattice potential and is denoted as $\phi_0$, so that the time-dependent potential reads

\begin{eqnarray}
V_{m,n}(t) &=& \kappa\ \textrm{cos} (m \pi /2  -\pi/4 ) \times \label{eq:mod} \\ \nonumber
&&\textrm{cos}(- n \pi / 2 + \omega_r t +\phi_{0} ) \\ \nonumber 
&+& \kappa\ \textrm{cos}(m \pi/2  +\pi/4 ) \times \\ \nonumber
&&\textrm{cos}(-n \pi/2 + \omega_b t+ \pi/2 +  \phi_{0} ) \ .
\end{eqnarray}

For resonant modulation $\omega_r=-\omega_b=\Delta/\hbar$, the total time-dependent Hamiltonian can be mapped onto an effective time-independent Hamiltonian with effective tunneling amplitudes $J^{\mathrm{eff}}_x\simeq J_x \kappa/(\sqrt{2}\Delta)$ and $J^{\mathrm{eff}}_y\simeq  J_y$ (see also Sect.~\ref{sect:effectiveHam}). The effective coupling along $x$ is complex with spatially-dependent phases $\phi_{m,n}=\phi_0 + \frac{\pi}{2} (m+n)$, which defines our experimental gauge and results in a flux of $\Phi=\phi_{m,n+1}-\phi_{m,n}=\pi/2$ per plaquette, aligned along the -$\hat{\mathbf{e}}_z-$direction. Note that in the main text and Sect.~\ref{sect:theorygauge} of the Supplementary Information we have chosen to describe our system using the Landau gauge $\phi_{m,n}=n \pi/2$ for the sake of simplicity.

\section{Laser-assisted tunneling on every other bond}

\begin{figure}[thb]
\begin{center}
\includegraphics[scale=1]{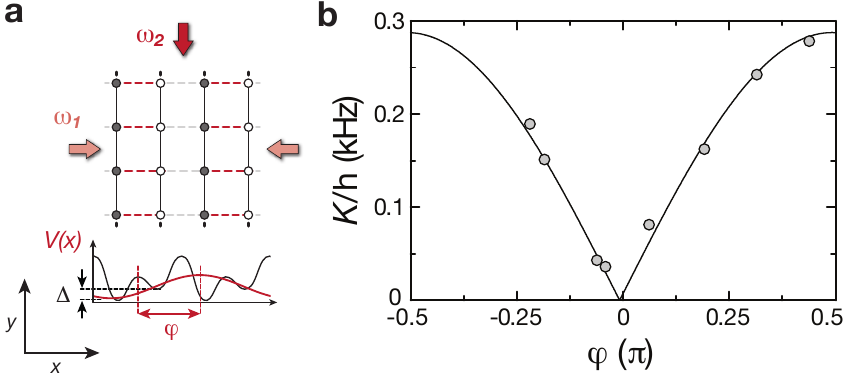}
\caption{Suppression of tunneling on every other bond \textbf{(a)} Schematic drawing of the experimental setup. Along $x$ a tilted double-well potential with energy offset $\Delta$ is used to inhibit tunneling. A pair of beams (red arrows) with frequency difference $\omega$ is then used to restore resonant tunneling. The corresponding local optical potential $V(x)$ is shown in red for $y=t=0$. The relative phase between the modulation and the underlying lattice is denoted as $\varphi$ \textbf{(b)}. The gray data points show the measured effective coupling strength induced by the modulation as a function of the relative phase $\varphi$. The solid line is a fit of eq.~(\ref{eq:drive}), where the amplitude and the phase offset are used as free fit parameters.}\label{fig:S2}
\end{center}
\end{figure}

Experimentally we studied the applicability of our flux-rectification method by measuring the suppression of tunneling on every other bond using only one pair of the beams mentioned above (see Sect.~\ref{sec:rectification}). The measurement was performed in isolated tilted double-well potentials (see Fig.~\ref{fig:S2}a), where tunneling on every other bond was suppressed due to a high potential barrier and tunneling within the double well was inhibited due to a potential offset between neighboring sites. We then restored resonant tunneling within the double-wells using only one pair of beams for the modulation and measured the effective tunneling amplitude as a function of $\varphi$, the phase relative to the underlying lattice. The effective tunnel coupling is proportional to the relative modulation amplitude between neighboring sites, which can be expressed as

\begin{equation}
V_{m+1,n}(t)-V_{m,n}(t)=\sqrt{2} \kappa | \textrm{sin}(\varphi) | \textrm{cos}(-n \pi/2 + \omega t) \label{eq:drive}
\end{equation}

\noindent where we have chosen the convention that for $\varphi=0$ the modulation along $x$ is in phase with the long lattice used to create the double-well potential and thus corresponds to a modulation that is in-phase on the two sites of the double-well and therefore cannot induce tunneling.

The experimental sequence started by loading a Bose-Einstein condensate of $^{87}$Rb atoms into a 3D optical lattice of depths $V_z=30(1)\,$Er$_z$, $V_y=30(1)\,$Er$_s$ and $V_{xL}=35(1)\,$Er$_L$, with Er$_{\alpha}=\hbar^2 k_{\alpha}^2/(2m)$,  $\alpha=\{s,L,z\}$, and $k_z=2\pi /\lambda_z$, with $\lambda_z=844\,$nm, $\lambda_s=767\,$nm and $\lambda_L=2\lambda_s$. After applying a filtering sequence where all double occupancies were removed, the short lattice along $x$ was ramped up to $V_{x}=7.0(2)\,$Er$_s$ within $20\,$ms. The phase of the superlattice was chosen such that a tilted double well with an energy offset of $\Delta/\hbar\approx 4.5\,$kHz was created where all atoms were located in the lower energy sites. Then we switched on the modulation instantaneously in order to induce resonant tunneling. The effective tunnel coupling was then determined by looking at the Rabi oscillations between the left and right wells of the double-well potentials as a function of the holdtime. The corresponding data is shown in Fig.~\ref{fig:S2}. From this we obtain an upper limit for the suppression of tunneling on every second bond of $J_{x,\textrm{min}}^{\textrm{eff}}/J_{x,\textrm{max}}^{\textrm{eff}}< 0.13$. A residual coupling is most likely due to imperfect reflection of the retro-reflected beam along $x$.

\section{The effective Harper-Hofstadter Hamiltonian}
\label{sect:effectiveHam}

The experimental setup used to rectify the artificial magnetic flux in a two-dimensional superlattice potential as described in Sect.~\ref{sec:rectification} gives rise to an explicitly time-dependent Hamiltonian, which can be separated as
\be
\hat H (t)=\hat H_0 + \hat V (t),
\ee 
where $\hat H_0$ describes the static components,  and $\hat V (t)$ is the time-dependent modulation. 

Considering a single-band tight-binding approximation, the static Hamiltonian $\hat H_0$ is taken in the form
\begin{align}
\hat{H}_0=&\hat T_x + \hat T_y + \hat V_{\text{conf}} + \hat U_{\text{int}} \label{eq:TB} \\
&+ \frac{\Delta}{2} \sum_{m,n} (-1)^{m} \hat n_{m,n} + \frac{\delta}{2} \sum_{m,n} \left [(-1)^{m} + (-1)^{n} \right ] \hat n_{m,n}. \notag
\end{align}
The first line includes the confining potential, the nearest-neighbor hopping and interaction terms,
\begin{align}
&\hat T_x= -J_x \sum_{m,n} \hat a_{m+1,n}^{\dagger} \hat a_{m,n} + \hat a_{m-1,n}^{\dagger} \hat a_{m,n} ,\notag \\
&\hat T_y= -J_y \sum_{m,n} \hat a_{m,n+1}^{\dagger} \hat a_{m,n} + \hat a_{m,n-1}^{\dagger} \hat a_{m,n} ,\notag \\
&\hat V_{\text{conf}} = \sum_{m,n} \hat n_{m,n} V_{\text{conf}}(m,n) , \notag \\
&\hat U_{\text{int}} = (U/2)  \sum_{m,n} \hat n_{m,n} \left (\hat n_{m,n} -1  \right ) , \notag
\end{align}
where $J_{x,y}$ denote the hopping matrix elements, $\hat a_{m,n}^{\dagger}$ creates a particle at lattice site $\bs x = (ma,na)$, $a$ is the lattice spacing, $(m,n)$ are integers and $U$ denotes the on-site interaction strength. The number operator is defined as $\hat n_{m,n} = \hat a_{m,n}^{\dagger} \hat a_{m,n}$, and the external harmonic trapping potential $V_{\text{conf}}(m,n) \propto (m^2 + n^2)$. The second line in eq.~(\ref{eq:TB}) describes the main staggered potential with energy offset $\Delta+\delta$ between neighboring sites, which inhibits tunneling along the $x$ direction and an auxiliary weak staggered potential with offset $\delta$ along $y$. The additional offset $\delta$ gives rise to a symmetric staggered detuning along both directions, with $\delta \ll \Delta$. 

In the tight-binding limit in which we work, the main effect of the additional pairs of beams (see Sect.~\ref{sec:rectification}) is a modulation of the on-site energies, giving a time-dependent contribution to the Hamiltonian of the form 
\begin{align}
\hat V (t)\!&=\! \kappa \sum_{m,n} \hat n_{m,n} \left \{  f_r(m) \cos [\omega t \!+\! g_r(n)] \right. \notag \\
&\left. \quad \quad \!+\!  f_b(m) \cos [\omega t \!+\! g_b(n)]     \right \} ,\notag \\
f_r(m) &= \cos (m \pi /2 -\pi/4), \ f_b(m)= \cos (m \pi /2 + \pi/4), \notag \\
g_r(n)&= \phi_0 - n \pi /2 , \ g_b(n)= n \pi /2 - \phi_0 - \pi/2 , \label{def_f_fun}
\end{align}
\noindent where $\phi_0$ is determined by the relative phase between the running-waves along $y$ and the underlying lattice potential, which is not controlled in the current experimental setup (see Sect.~\ref{sec:rectification}). In order to restore the tunneling along the $x$ direction, we fix the modulation frequency $\omega$ so as to satisfy the resonance condition $\hbar \omega=\Delta$. 

The time-evolution of the system is ruled by the Schr\"odinger equation $i \hbar \partial_t \psi = \hat H(t) \psi$. Let us consider the unitary transformation
\begin{align}
&\psi = \hat R(t) \tilde \psi = \exp \left (\! -i \frac{\omega}{2} \hat W t \right ) \tilde \psi,\label{transf_hof}\\
&\hat W = \sum_{m,n} (-1)^{m} \hat n_{m,n}.\notag
\end{align}
The transformed state satisfies the Schr\"odinger equation $i \hbar \partial_t \tilde \psi = \tilde H(t) \tilde \psi$, with the modified Hamiltonian
\be
\tilde H (t) = \tilde H_0 + \hat V^{(+1)}e^{i \omega t} + \hat V^{(-1)}e^{-i \omega t},\label{ham_dec}
\ee
where
\begin{eqnarray}
\tilde H_0 &=& \hat T_y + \hat V_{\text{conf}} +  \hat U_{\text{int}} + \frac{\delta}{2} \sum_{m,n} \left [(-1)^{m} + (-1)^{n} \right ] \hat n_{m,n} \label{def_eff_operator} \nonumber \\
\hat V^{(+1)} &=& \frac{\kappa}{2}\sum_{m,n} \hat n_{m,n}\,  \mathfrak{g} (m,n) \\
&&\quad \quad \ -J_x\! \sum_{m \text{odd}, n} \! \hat a_{m+1,n}^{\dagger} \hat a_{m,n} \!+\! \hat a_{m-1}^{\dagger} \hat a_{m,n}, \notag \\
\hat V^{(-1)} &=& \frac{\kappa}{2}\sum_{m,n} \hat n_{m,n}\,  \mathfrak{g}^* (m,n) \nonumber \\
&&\quad \quad \ -\! J_x\! \sum_{m \text{even}, n} \! \hat a_{m+1,n}^{\dagger} \hat a_{m,n} \!+\! \hat a_{m-1}^{\dagger} \hat a_{m,n},\notag
\end{eqnarray}
where
\be
\mathfrak{g} (m,n) = f_r(m) e^{i  g_r (n) } +  f_b(m) e^{i g_b (n) }.\label{def_f_g}
\ee

We describe the time-evolution of the system by partitioning the evolution operator as
\begin{align}
\hat U(t) =e^{-i\hat K(t)}e^{- i t \hat H_{\rm eff}/\hbar}e^{i\hat K(0)},
\end{align}
where the effective Hamiltonian $\hat H_{\rm eff}$ describes the long-time dynamics, and where the operator $\hat K(t)$ captures the micro-motion, see Refs.~\cite{GoldmanDalibard:PRX,Rahav}. Note that the initial kick $e^{i\hat K(0)}$, which depends on the initial phase of the modulation, is inhibited in the experiment by launching the modulation adiabatically. Following Ref.~\cite{GoldmanDalibard:PRX}, we find that the effective Hamiltonian associated with the general single-harmonic time-dependent Hamiltonian in eq.~\eqref{ham_dec} is given by
\begin{align}
&\hat H_{\rm eff}=\tilde H_0 + \frac{1}{\hbar \omega }  [\hat V^{(+1)} , \hat V^{(-1)} ] \label{effective_ham_one} \\
&+ \frac{1}{2 (\hbar \omega)^2} \left ( [[\hat V^{(+1)},\tilde H_0],\hat V^{(-1)} ] + [[\hat V^{(-1)},\tilde H_0],\hat V^{(+1)} ] \right )  \notag \\
& + \mathcal{O} (1/\omega^3), \notag 
\end{align}
where we considered a perturbative expansion in powers of $(1/\omega)$. To be explicit, we introduce a small dimensionless quantity $\Omega_{\text{eff}}/\omega \ll 1$, where $\Omega_{\text{eff}}$ is a typical frequency associated with the effective Hamiltonian (see below). In the following, we will identify $\Omega_{\text{eff}}$ with the cyclotron frequency associated with the Harper-Hofstadter Hamiltonian \eqref{final_ham}, $\Omega_{\text{eff}}=B/m^*$, where $B=\Phi \hbar /a^2$ is the effective magnetic field, $\Phi$ is the related flux per plaquette and $m^*=\hbar^2/(2 J a^2)$ is the effective mass with $J=J_x^{\text{eff}}=J_y$. In the present experimental scheme, the flux will be found to be $\Phi=2 \pi (1/4)$, see eq.~\eqref{final_ham} below, so that we obtain $\Omega_{\text{eff}}= \pi J / \hbar$; hence, the driving should satisfy the high-frequency condition $\hbar \omega = \Delta \gg J$, which is indeed the case in the experiment. 

We now evaluate the effective Hamiltonian in eq.~\eqref{effective_ham_one}, up to second-order  in $(1/\omega)$,
\be
\hat H_{\rm eff} =  \hat H_{\rm eff}^{(0)} + \hat H_{\rm eff}^{(1)} + \hat H_{\rm eff}^{(2)} + \mathcal{O} (1/\omega^3)
\ee
using the specific operators defined in eq.~\eqref{def_eff_operator}.

\subsection*{The zeroth-order terms}

The zeroth order contribution to the effective Hamiltonian is given by the static terms
\begin{eqnarray}
\hat H_{\rm eff}^{(0)}\!=\!\tilde H_0 &=& \hat T_y + \hat V_{\text{conf}} \\
&&+ \hat U_{\text{int}} + \frac{\delta}{2} \sum_{m,n} \left [(-1)^{m} \!+\! (-1)^{n} \right ] \hat n_{m,n}, \nonumber
\end{eqnarray}
which signals the absence of tunneling along the $x$ direction at the lowest order of the calculations. 

\subsection*{The first-order terms}

The first-order contributions to the effective Hamiltonian are given by
\begin{eqnarray}
\hat H_{\text{eff}}^{(1)} &=&  \frac{1}{\hbar \omega }  [\hat V^{(+1)} , \hat V^{(-1)} ]  \label{first_corr}  \\
&=& J_x^{\text{eff}} \sum_{m,n}  \hat a_{m+1,n}^{\dagger} \hat a_{m,n} e^{i \phi_{m,n}}  +  \text{h.c.}, \notag \\
J_x^{\text{eff}}  &=& J_x \frac{ \kappa}{\sqrt{2} \hbar \omega}, \quad \phi_{m,n} =\left [ \pi/2 (m+n) - \phi_0   \right ], \nonumber
\end{eqnarray}
hence simultaneously restoring the hopping along the $x$ direction and generating space-dependent Peierls phases $\phi_{m,n}$.  In the experiment, the parameters are chosen such that the hopping is approximately homogeneous along both spatial directions, yielding $J_x^{\text{eff}}  \approx J_y \approx 75 \text{Hz} \times h$. Consequently, the first-order contributions are as important as the zero-th order terms. \\

In summary, the first-order effective Hamiltonian reproduces the Harper-Hofstadter model~\cite{Hofstadter} with a uniform flux $\Phi= 2 \pi (1/4)=\pi/2$ per plaquette,
\begin{align}
&\hat H_{\rm eff} =  \hat H_{\rm eff}^{(0)} + \hat H_{\rm eff}^{(1)} , \label{final_ham} \\
&\quad \, \, \, \, \, =   J_x^{\text{eff}} \sum_{m,n} \left \{ \hat a_{m+1,n}^{\dagger} \hat a_{m,n} e^{i \left [ \pi/2 (m+n) - \phi_0   \right ]}  +  \text{h.c.} \right \} \notag \\
& \quad \, \, \, \, \, \,  -J_y \sum_{m,n} \left\{ \hat a_{m,n+1}^{\dagger} \hat a_{m,n} + \hat a_{m,n-1}^{\dagger} \hat a_{m,n} \right \}+ \hat V_{\text{extra}}, \notag\\
& \hat V_{\text{extra}} = \frac{\delta}{2} \sum_{m,n} \left [(-1)^{m} \!+\! (-1)^{n} \right ] \hat n_{m,n} + \hat V_{\text{conf}}+  \hat U_{\text{int}}  \notag .
\end{align}
At this order of the calculations, all additional effects [i.e. the static staggered potential detuning, the confinement and interactions] are assembled in $\hat V_{\text{extra}}$. Note that for the Hamiltonian given in eq.~(1) of the main text and the discussion of the energy spectrum in Sect.~\ref{sect:theorygauge} of the Supplementary Information, we have chosen to describe our system in the Landau gauge $\phi_{m,n}=n \pi/2$ for the sake of simplicity.

\subsection*{The second-order terms}

The second-order contributions lead to four distinct corrections:
\begin{align}
\hat H_{\text{eff}}^{(2)}&=\frac{1}{2 (\hbar \omega)^2} \left ( [[\hat V^{(+1)},\tilde H_0],\hat V^{(-1)} ] + [[\hat V^{(-1)},\tilde H_0],\hat V^{(+1)} ] \right )\notag \\
&= \mathcal{C}_1 + \mathcal{C}_2 +  \mathcal{C}_3 + \mathcal{C}_4 .  \notag
\end{align}
In the following, we omit the high-order contributions stemming from the on-site interaction term $\hat U_{\text{int}}$, which lead to negligible delocalized interaction terms.

The first and main correction is a space-dependent renormalization of the hopping along the $y$ direction:
\begin{align}
&\mathcal{C}_1 = -J_y \left (\frac{\kappa}{2 \hbar \omega} \right )^2 \sum_{m,n} \mu_{m,n}  \hat a_{m,n+1}^{\dagger} \hat a_{m,n}  + \text{h.c.} \notag \\
&\mu_{m,n} = - 2 + 2 (-1)^{m+n} \cos (2 \phi_0). \label{inhom_y_corr}
\end{align}
Thus, taking into account this second-order correction, we find that the hopping amplitude is potentially inhomogeneous along the $y$ direction, and that it ranges between $J_y^{\text{eff}} = J_y $ and $J_y^{\text{eff}}=J_y[ 1 - (\kappa/(\hbar\omega))^2 ]$. As discussed in the next Section, the parameter $\kappa/(\hbar\omega) \approx 0.58$ in the experiment, so that this effect cannot be safely neglected. 

The three other corrections are very weak, and thus, they can be neglected in the regimes explored by the experiment. To be explicit, the second correction is a weak next-nearest-neighbor hopping term of the form
\begin{align}
\mathcal{C}_2 =  \frac{\kappa J_x J_y}{(\hbar \omega)^2}  \sum_{m,n}& e^{i \theta_1} \hat a_{m+1,n+1}^{\dagger} \hat a_{m,n} + e^{i \theta_2} \hat a_{m-1,n+1}^{\dagger} \hat a_{m,n}  \notag \\
& + e^{i \theta_3} \hat a_{m+1,n-1}^{\dagger} \hat a_{m,n} + e^{i \theta_4} \hat a_{m-1,n-1}^{\dagger} \hat a_{m,n} \notag .
\end{align}
The third effect is an even weaker correction to the nearest-neighbour hopping term along $x$,
\begin{eqnarray}
\mathcal{C}_3 = \frac{\kappa J_x}{2 (\hbar \omega)^2} \sum_{m \text{ odd} , n } &&\lambda_{m,n} \hat a_{m+1,n}^{\dagger} \hat a_{m,n} \nonumber \\
&&+ \tilde \lambda_{m,n} \hat a_{m-1,n}^{\dagger} \hat a_{m,n}  + \text{h.c.} \nonumber 
\end{eqnarray}
where $\lambda_{m,n}$ and $\tilde \lambda_{m,n}$ depend linearly on the additional (weak) potentials in $\hat V_{\text{extra}}$, see Eq. \eqref{final_ham}. The fourth effect $\mathcal{C}_4$ is a negligible  next-nearest-neighbor hopping term proportional to $J_x J_y / (\hbar \omega)^2 \ll 1$.\\

\subsection*{Summary}

In conclusion, in the parameters regime considered in the experiment, we find that the dynamics should be well described by the first-order effective (Harper-Hofstadter) Hamiltonian in eq.~\eqref{final_ham}, including the second-order corrections $\mathcal{C}_1$ presented in eq.~\eqref{inhom_y_corr}:
\begin{align}
\hat H_{\rm eff}  &=   -J \sum_{m,n} \bigl \{ \hat a_{m+1,n}^{\dagger} \hat a_{m,n} e^{i \left [ \pi/2 (m+n) - \phi_0   \right ]}  +  \text{h.c.}  \notag \\
& \qquad \qquad + (1+f_{m,n}) \hat a_{m,n+1}^{\dagger} \hat a_{m,n} + \text{h.c.}\bigr \} \notag\\
&+\frac{\delta}{2} \sum_{m,n} \left [(-1)^{m} \!+\! (-1)^{n} \right ] \hat n_{m,n} + \hat V_{\text{conf}} +  \hat U_{\text{int}} \label{sim_ham_new} ,
\end{align}
where we introduced the correction to the hopping along $y$
\be
f_{m,n} = - \frac{1}{2} \left ( \frac{\kappa}{\hbar \omega} \right )^2 \left \{ 1 - (-1)^{m+n} \cos (2 \phi_0) \right \},\label{inhom_y_corr_bis_new}
\ee
and where we set $J=J_y$. Moreover, in this regime the static linear gradient used to generate the Hall drift can be simply added according to 

\begin{equation}
\hat{H}_{\text{eff}} \rightarrow \hat{H}_{\text{eff}}-Fa  \sum_{m,n} n \hat{n}_{m,n}. \nonumber 
\end{equation}

\section{Energy spectrum and magnetic unit cell}
\label{sect:theorygauge}

The energy spectrum of non-interacting particles in a periodic potential exposed to an external magnetic field is described by the well-known Hofstadter butterfly~\cite{Hofstadter}. This structure can be understood, starting from a simple tight-binding description. In the absence of the magnetic field, the tight-binding approximation leads to a single energy band, $E =-2 J [\cos (k_x a) + \cos (k_y a)]$, where $J$ is the tunneling matrix element between neighboring sites. Adding the magnetic field, through the introduction of Peierls phases~\cite{Hofstadter}, leads to a fractionalization of the tight-binding band into several subbands: in particular, when the magnetic flux per plaquette is given by $\Phi = 2 \pi \alpha=2 \pi (p/q)$, with $p, q$ integers, the band splits into $q$ subbands. In our experimental setup $\alpha=1/4$, so that the energy spectrum is constituted of four subbands. Besides, the magnetic flux effectively extends the standard unit cell of the lattice into a ``magnetic unit cell", which in this case is constituted of $q=4$ lattice sites. In the following, and for the sake of simplicity, we have chosen to describe the system using the Landau gauge together with a square [``symmetric"] 4-site unit cell, see the gray shaded area in Fig.~1a of the main text and below. 

We start with the Schr\"odinger equation associated with the Harper-Hofstadter model~\cite{Hofstadter},
\begin{eqnarray}
\varepsilon \Psi_{m,n} &=& e^{\ii n \pi/2} \Psi_{m+1,n} \nonumber \\
&&+ e^{-\ii n \pi/2} \Psi_{m-1,n} + \Psi_{m,n+1} + \Psi_{m,n-1},\notag
\end{eqnarray}
which describes hopping on the square lattice in the presence of a magnetic flux $\Phi=\pi/2 $ per plaquette. Here $\varepsilon= - E/J$, $(m,n)$ are integers labeling the lattice sites, and for now, we considered that the hopping is homogeneous along both directions (i.e. $J_x^{\text{eff}}=J_y=J$  in terms of the experimental parameters discussed in the previous section). To solve this equation, we make the following ansatz for the wave function:
\begin{equation}
\Psi_{m,n} = e^{\ii k_x m} e^{\ii k_y n} 
\begin{cases} 
\psi_A ,&\textrm{for} \ m,n  \ \textrm{even}\\ 
\psi_B  \ e^{\ii n \pi/2 } & \textrm{for} \ m  \ \textrm{odd},  \ n  \ \textrm{even}\\
\psi_C  &  \textrm{for} \ m  \ \textrm{even},  \ n  \ \textrm{odd}\\
\psi_D \ e^{\ii n \pi/2} &\textrm{for} \ m,n  \ \textrm{odd}
\end{cases} \nonumber
\end{equation}
\noindent where $k_x$, $k_y$ are defined within the first magnetic Brillouin zone ($k_x \in [-\pi/(2a),\pi/(2a)[$, $k_y \in [-\pi/(2a),\pi/(2a)[$). Inserting this ansatz into the Schr\"odinger equation we obtain the following $4 \times 4$ eigenvalue equation
\begin{equation}
\hat{H} 
\begin{pmatrix}
\psi_A \\
\psi_B \\
\psi_C \\
\psi_D
\end{pmatrix} 
= E(\mathbf{k})
\begin{pmatrix}
\psi_A \\
\psi_B \\
\psi_C \\
\psi_D
\end{pmatrix} ,
\end{equation}

\noindent with

\begin{equation}
\hat{H} = -2 J
\begin{pmatrix}
0 &  \cos k_x &   \cos k_y & 0 \\
 \cos k_x & 0 & 0 & - \sin k_y \\
 \cos k_y & 0 & 0 & - i  \sin k_x \\
0 & - \sin k_y & i  \sin k_x & 0
\end{pmatrix} ,
\end{equation}
where we set $a=1$.

Adding the staggered potential detuning (see main text and Eq. \eqref{final_ham} above),
\begin{align}
\frac{\delta}{2} \sum_{m,n} \left [(-1)^{m} + (-1)^{n} \right ] \hat n_{m,n}, \notag
\end{align}
where $\hat n_{m,n}$ is the particle number operator at lattice site $(m,n)$, leads to the modified Hamiltonian matrix
\begin{equation}
\hat{H} \rightarrow \hat{H} = -2 J
\begin{pmatrix}
(-\delta/2J) &  \cos k_x &   \cos k_y & 0 \\
 \cos k_x & 0 & 0 & - \sin k_y \\
 \cos k_y & 0 & 0 & - i  \sin k_x \\
0 & - \sin k_y & i  \sin k_x & (\delta/2J)
\end{pmatrix} .\label{ideal_harper_stag}
\end{equation}

\begin{figure}[thb]
\begin{center}
\includegraphics{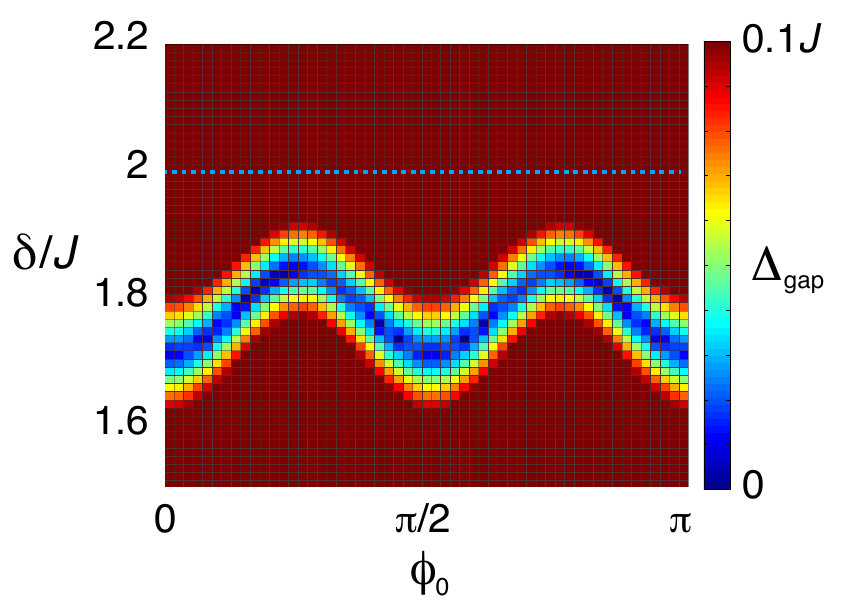}
\caption{Gap closing point, separating the topological and non-topological regimes, as a function of the staggered potential detuning $\delta$ and the relative phase $\phi_0$ for $\kappa /( \hbar \omega ) = 0.58$. Here $J_x^{\text{eff}}=J_y=J$, where $J_y$ is the bare tunneling amplitude along $y$. The blue dotted line shows the transition point for the homogeneous Harper-Hofstadter model in eq.~\eqref{ideal_harper_stag}, i.e. $J_x^{\text{eff}}=J_y^{\text{eff}}=J$.}\label{phase_dia}
\end{center}
\end{figure}

Finally, to build the complete effective Hamiltonian, we consider the contribution of the inhomogeneous hopping along the $y$ direction, which stems from the second-order corrections in eq.~\eqref{inhom_y_corr}, see Section~\ref{sect:effectiveHam}. The total hopping term along the $y$ direction [including the zero-th order ``bare" hopping and the second-order corrections] is written as [eq.~\eqref{sim_ham_new}]
\begin{eqnarray}
\hat T_y &=& -J \sum_{m,n} (1 + f_{m,n})  \hat a_{m,n+1}^{\dagger} \hat a_{m,n}  \notag \\
&& \quad \quad \quad + (1 + f_{m,n-1})  \hat a_{m,n-1}^{\dagger} \hat a_{m,n}, \notag \\
f_{m,n} &=& - \frac{1}{2} \left ( \frac{\kappa}{\hbar \omega} \right )^2 \left \{ 1 - (-1)^{m+n} \cos (2 \phi_0) \right \},\label{inhom_y_corr_bis}
\end{eqnarray}
where we set $J=J_y$ is the bare hopping along the $y$ direction (and we remind that $J_x^{\text{eff}}=J_y=J$). Taking these corrections into account (see Section \ref{sect:effectiveHam}) leads to the final effective Hamiltonian matrix
\begin{widetext}
\begin{align}
&\hat{H}_{\text{eff}} \!=\! -2 J
\begin{pmatrix}
(-\delta/2J) &  \cos k_x &   \cos k_y + h_1& 0 \\
 \cos k_x & 0 & 0 & - \sin k_y + h_2^{*} \\
 \cos k_y + h_1^* & 0 & 0 & - i  \sin k_x \\
0 & - \sin k_y +h_2 & i  \sin k_x & (\delta/2J)
\end{pmatrix}  \notag \\
& h_1 = - \frac{1}{2} \left ( \frac{\kappa}{\hbar \omega} \right )^2 \left [ \cos k_y - i \cos (2 \phi_0) \sin k_y    \right ], \quad 
h_2 =  \frac{1}{2} \left ( \frac{\kappa}{\hbar \omega} \right )^2 \left [ \sin k_y + i \cos (2 \phi_0) \cos k_y    \right ]. \label{good_ham_bulk}
\end{align}
\end{widetext}
In the experiment, $\kappa /( \hbar \omega) \approx 0.58$, so that the effects related to the inhomogeneous hopping along the $y$ direction cannot be neglected [i.e. the hopping can potentially be reduced locally by $30 \%$]. We illustrate this effect by computing the gap closing point, separating the topological and non-topological regimes and driven by the staggered potential detuning $\delta$, for $\kappa /( \hbar \omega )= 0.58$ and $\phi_0 \in [0 , \pi]$. As shown in Fig.~\ref{phase_dia}, the ideal transition point $\delta=2 J$, corresponding to the homogeneous Harper-Hofstadter model \eqref{ideal_harper_stag} [$J_x^{\text{eff}}=J_y^{\text{eff}}$], is shifted and oscillates as a function of the relative phase $\phi_0$. We find that these effects are of the order of experimental uncertainties, indicating that higher-order effects can be safely neglected in the analysis. In particular, omitting the effects due to the external trap and inter-particle interactions, we find that the setup and its phase transitions are well described by the $4 \times 4$ Hamiltonian matrix in eq.~\eqref{good_ham_bulk}.

\section{The anomalous velocity, the Hall drift and the effective Chern number}\label{section:anomalous}

\subsection*{General discussion}

We start the discussion by considering a square lattice subjected to a general flux per plaquette $\alpha=p/q$ [in units of the flux quantum, i.e. $\Phi= 2 \pi \alpha$]. Each ``magnetic" unit cell of the lattice contains $q$ lattice sites, and the energy spectrum splits into $q$ subbands. In the case where these bands are well separated, the number of states available in each subband is $N_{\text{state}} = N_x N_y$, where $N_{x,y}$ are the numbers of magnetic unit cells along each spatial direction [there are then $q N_xN_y$ lattice sites in the system]. \\

The $q$ bands are labelled as $\varepsilon_{\mu} (k_x,k_y)$, where ${\mu}=1, \dots, q$, and the quasi-momenta [Bloch parameters] take the values
\begin{align}
&k_x = \frac{2 \pi n_x}{L_x} , \quad k_y = \frac{2 \pi n_y}{L_y} , \quad n_{x,y}=1, \dots , N_{x,y} ,\\
&\Delta_{k_x}=\frac{2 \pi}{q N_x a} , \quad \Delta_{k_y}=\frac{2 \pi}{N_y a},
\end{align}
where we have chosen the magnetic unit cell to have length $q a$  along the $x$ direction, such that the overall system has lengths $L_x=q N_x a$ and $L_y=N_y a$ (In the main text, for $q=4$, the magnetic unit cell is chosen to have size $2a \times 2a$, such that $L_{x,y} = 2N_{x,y}a$ and the resulting magnetic Brillouin zone is symmetric with respect to $k_x$ and $k_y$). The first ``magnetic" Brillouin zone (FBZ) is defined by  
\be
k_x \in (-\pi/qa, \pi/qa] , \quad k_y \in (-\pi/a ,  \pi/a] . \label{FBZ}
\ee
We act on the particles with a force $\bs F = F \hat{\mathbf{e}}_y$, aligned along $y$, and we focus the following analysis on the velocity $v=v^x$ defined in the direction perpendicular to the force. The average velocity in a state $\vert u_{\mu}(k_x,k_y) \rangle$ associated with the energy $\varepsilon_{\mu} (k_x,k_y)$ is given by~\cite{Xiao}
\be 
v_{\mu}(k_x,k_y) = v^{\text{band}}_{\mu} - \frac{F}{\hbar}  \Omega_{\mu} (\bs k),\label{ano_vel}
\ee
where $\Omega_{\mu} (\bs k)$ is the Berry curvature of the ${\mu}$-th band,
\be
\Omega_{\mu} (\bs k)  = i \left ( \langle \partial_{k_x} u_{\mu} \vert \partial_{k_y} u_{\mu} \rangle - \langle \partial_{k_y} u_{\mu} \vert \partial_{k_x} u_{\mu} \rangle \right ), \label{Berrycurv}
\ee 
and $v^{\text{band}}_{\mu}= (1/\hbar) \partial_{k_x} \varepsilon_{\mu}$ is the standard band velocity along $x$. The contribution due to the Berry's curvature in eq.~\eqref{ano_vel} is generally referred to as the ``anomalous velocity", see Ref.~\cite{Xiao}.

We now populate each band $\varepsilon_{\mu}$ with  $N^{({\mu})}$ particles. We assume that these particles uniformly distribute themselves over the entire band (this assumption is validated experimentally through the band-mapping technique, see also Sect.~\ref{sect:bandmapping}). We write the total number of particles as
\be
N_{\text{tot}} = N^{(1)} + N^{(2)} + \dots + N^{({\mu})}+ \dots + N^{(q)}. 
\ee
The mean number of particles in a state $\vert u_{\mu}(k_x,k_y) \rangle$ is then given by
\be
\rho^{({\mu})}(\bs k) = \frac{N^{({\mu})}}{N_xN_y} = \rho^{({\mu})}.
\ee

The mean velocity per particle is given by
\begin{align}
\langle v \rangle &= \frac{1}{N_{\text{tot}}} \sum_{\mu} \sum_{\bs k} \rho^{({\mu})}(\bs k) \, v^{({\mu})}(\bs k), \notag \\
&= \frac{1}{N_{\text{tot}}} \sum_{\mu} \rho^{({\mu})} \sum_{\bs k} \left ( - \frac{F}{\hbar}  \Omega_{\mu} (\bs k)    \right ), \notag \\
&= \frac{1}{N_{\text{tot}}} \sum_{\mu} \rho^{({\mu})}  \left ( - \frac{2 \pi}{\hbar} F \Delta_{k_x}^{-1} \Delta_{k_y}^{-1}  \right ) \notag \\
&\qquad \quad \qquad \qquad \underbrace{\times \frac{1}{2 \pi}\sum_{\bs k} \Omega_{\mu} (\bs k) \Delta_{k_x} \Delta_{k_y}}_{\rightarrow \nu_{\mu}} ,\notag \\ 
&= \frac{1}{N_{\text{tot}}} \sum_{\mu} \rho^{({\mu})}  \left ( - \frac{F q N_x N_y a^2}{h} \right )  \nu_{\mu} ,\notag \\ 
&= \frac{1}{N_{\text{tot}}} \sum_{\mu} N^{({\mu})}  \left ( - \frac{F q  a^2}{h} \right )  \nu_{\mu}, \notag 
 \end{align}
 where we introduced the Chern number $\nu_{\mu}$ of the ${\mu}$-th band~\cite{Thouless,Simon,Kohmoto}
\begin{align}
&\nu_{\mu}=\frac{1}{2 \pi} \int_{\text{FBZ}} \Omega_{\mu} (\bs k) \text{d}^2 k.\label{chern}
\end{align}
Here, the velocity has no contribution from the band velocity $\partial_{k_x} \varepsilon_{\mu}$, which vanishes by symmetry when the band is uniformly filled~\cite{Xiao,Price,Dauphin}. 
  
Introducing the band filling factor $\eta_{{\mu}}$, we finally find the mean velocity per particle
\be
\langle v \rangle=  - \frac{F q  a^2}{h}  \sum_{\mu} \eta_{{\mu}}    \nu_{\mu} , \quad  \eta_{{\mu}}=N^{({\mu})} / N_{\text{tot}}.  \notag 
\ee
For $\eta_{{\mu}}(t)=\eta_{{\mu}}^0$ constant in time, the center-of-mass displacement along the $x$ direction is thus given by
\be
 x (t) = x(t_0)  - \frac{F t q  a^2}{2 \pi \hbar}  \sum_{\mu} \eta_{{\mu}}^0    \nu_{\mu}.  
\ee
We stress that we assumed that the force $F$ is weak enough so that the band populations remain constant during the motion.\\

If only the lowest band is filled, i.e. $\eta_{1}^0=1$ and $\eta_{{\mu}>1}^0=0$, we find
\be
 x (t) = x(t_0)  - \frac{F t q  a^2}{2 \pi \hbar} \nu_1,  
\ee
where $\nu_1$ is the Chern number of the filled (lowest) band~\cite{Dauphin}.\\

\subsection*{Energy spectrum displaying band touching points: the case $\Phi=1/4$}

\begin{figure}[t!]
\begin{center}
\includegraphics{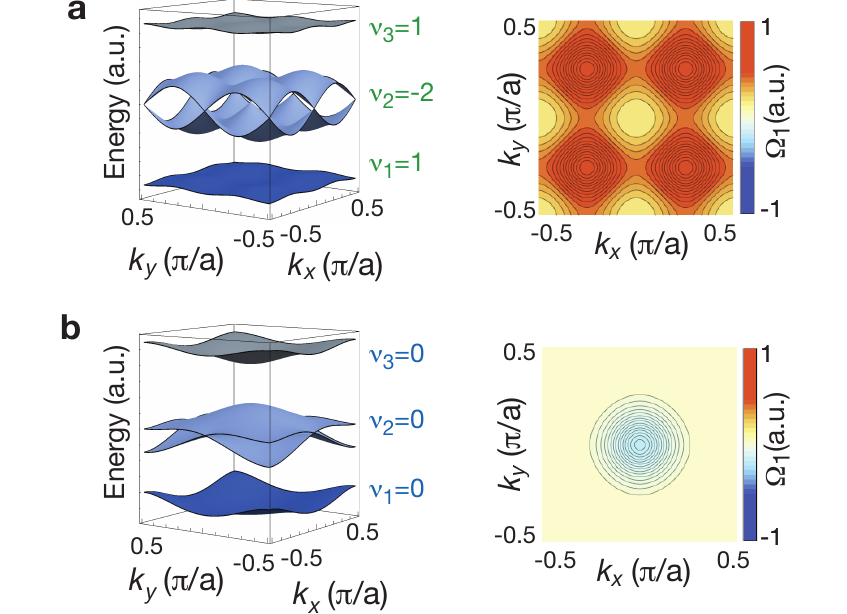}
\caption{Energy spectrum, Chern number distribution and Berry curvature of the lowest band for the Harper-Hofstadter Hamiltonian with flux $\Phi=\pi/2$. \textbf{a} the topological case $\delta=0$, and \textbf{b} the topologically trivial case triggered by the staggered potential detuning, here illustrated for $\delta=4J$. In contrast to the topologically trivial case, the Berry curvature has only positive contributions in the topological regime, leading to a non-zero Chern number $\nu_1=+1$.}\label{fig:berry_curvature}
\end{center}
\end{figure}

When the flux is $\Phi=\pi/2$ [i.e. $\alpha=1/4$ and $q=4$], as it is the case in the experiment [see eq.~\eqref{final_ham}], the energy spectrum only displays \emph{three well separated bands} (see Fig.~\ref{fig:berry_curvature}a and Fig.2 in the main text). The central ``super-band" consists of two touching subbands~\cite{Kohmoto}: $\varepsilon_2$ and $\varepsilon_{3}$. Since the Chern number is only well defined for isolated bands~\cite{Thouless,Kohmoto}, it is important to consider the new following labeling of bands:
\be
E_1 = \varepsilon_1 , \quad E_2 = `` \varepsilon_2 + \varepsilon_3" , E_3 = \varepsilon_4.
\ee
The Chern numbers $\nu_{1,2,3}$ associated with these well-separated bands have the values $\{ 1 , -2 , 1 \}$, see Ref.~\cite{Kohmoto}. \\

We now fill each band $E_{\mu}$ with  $N^{({\mu})}$ particles, and ${\mu}=1,2,3$. As before, we assume that these particles uniformly distribute themselves over each band. We write the total number of particles as
\be
N_{\text{tot}} = N^{(1)} + N^{(2)} +N^{(3)}. 
\ee
The mean number of particles in a state of the lowest or upper bands are
\be
\rho^{(1)}=\frac{N^{(1)}}{N_x N_y} , \quad \rho^{(3)}=\frac{N^{(3)}}{N_x N_y} , 
\ee
whereas in the central band, the mean number of particles is
\be
\rho^{(2)}=\frac{N^{(2)}}{ 2 N_x N_y} ,  
\ee
since the second ``super-band" contains $2 N_x N_y$ states. The mean velocity per particle is then obtained as in the previous Section, 
\begin{align}
&\langle v \rangle  = \frac{1}{N_{\text{tot}}} \sum_{{\mu}=1,2,3} \rho^{({\mu})} \sum_{\bs k} \left ( - \frac{F}{\hbar}  \Omega_{\mu} (\bs k)    \right ), \notag \\
& = - \frac{F}{\hbar N_{\text{tot}}} \left [ \rho^{(1)} \sum_{\bs k}  \Omega_1 (\bs k) + \rho^{(2)} \sum_{\bs k}  \Omega_2 (\bs k) \right. \notag \\
&\ \  \qquad \qquad \left. + \rho^{(3)} \sum_{\bs k}  \Omega_3 (\bs k) \right ], \notag \\
&=- \frac{4 F   a^2}{h N_{\text{tot}}}  \left [ N^{(1)} \nu_1 + \frac{N^{(2)}}{2} \nu_2 + N^{(3)} \nu_3  \right ] ,\notag  \\
&=- \frac{4 F   a^2}{h}  \left [ \eta_{1} \nu_1 + \frac{\eta_{2}}{2} \nu_2 + \eta_{3} \nu_3 \right ] ,\notag 
 \end{align}
where we again introduced the band filling factor $\eta_{{\mu}}=N^{({\mu})} / N_{\text{tot}}$, and the Chern numbers $\nu_{\mu}$ of the three separated bands, ${\mu}=1,2,3$. Accordingly, for $\eta_{{\mu}}(t)=\eta_{{\mu}}^0$ constant in time, the center-of-mass displacement is given by
\begin{align}
 x(t)  = x(t_0) - \frac{2 F t  a^2}{\pi \hbar}  \left [  \eta_{1}^0 \nu_1 + \frac{\eta_{2}^0}{2} \nu_2 + \eta_{3}^0 \nu_3 \right ] , \label{com-dis}
\end{align}
so that measuring  the initial fillings $\eta_{1,2,3}^0$ give access to the Chern number $ \nu_1$, using symmetry arguments. Indeed, in the next paragraph we show that $ \nu_1= \nu_3$, which is a direct consequence of the particle-hole symmetry inherent to the Harper-Hofstadter model~\cite{Hofstadter,Kohmoto}. Using this symmetry, together with the fact that the total tight-binding band carries a zero Chern number, i.e. $\sum_j \nu_j=0$, yields
\be
\nu_2= -2 \nu_1 , \quad \nu_3= \nu_1,
\label{eq:CNrelation}
\ee 
so that eq.~\eqref{com-dis} becomes
\begin{align}
 x(t)  = x(t_0) - \frac{2 F t  a^2}{\pi \hbar} \nu_1 \left [  \eta_{1}^0  - \eta_{2}^0+ \eta_{3}^0 \right ] . \label{eq:com-dis_0}
\end{align}
We can rewrite the latter result as
\begin{align}
& x(t) = x(t_0) - \frac{2 F t  a^2}{\pi \hbar} \nu_1^{\text{eff}} \\
&\nu_1^{\text{eff}} = \nu_1 \gamma_0 , \quad \gamma_0=\left [  \eta_{1}^0  - \eta_{2}^0+ \eta_{3}^0 \right ] ,
\end{align}
where $\nu_1^{\text{eff}}$ denotes the effective Chern number, which deviates from the ideal and quantized value $\nu_1$  when higher bands are initially populated [i.e. $\eta_{1}^0 < 1$].

When the populations vary in time, as it is the case in the experiment, the center-of-mass follows the equations of motion
\begin{align}
 x(t)  = x(t_0) - \frac{2 F a^2}{\pi \hbar} \nu_1 \int_0^t \eta_{1}(t')  - \eta_{2}(t')+ \eta_{3}(t') \text{d} t' . \label{com-dis_2}
\end{align}
Hence, measuring the populations $\gamma (t)=\eta_{1}(t)  - \eta_{2}(t)+ \eta_{3}(t)$ together with the COM displacement $x(t)$ allows to evaluate the Chern number $\nu_1$, based on long-time dynamics. \\

Finally, we note that an alternative analysis can be performed without invoking the particle-hole symmetry leading to eqs.~(\ref{eq:CNrelation})-(\ref{eq:com-dis_0}). Indeed, by imposing $\sum_j \nu_j=0$ only, the Hall deflection $x(t)$ can be fitted using the more general equation
\begin{equation}
\begin{split}
 x(t)  = x(t_0) - \frac{2 F a^2}{\pi \hbar} \biggl (& \nu_1 \int_0^t \eta_{1}(t')  - \eta_{3}(t') \text{d}t'  \\
  + &\nu_2 \int_0^t \frac{\eta_{2}(t')}{2}  - \eta_{3}(t') \text{d}t' \biggr ), \label{com-dis_3}
 \end{split}
\end{equation}
where the two fitting parameters are the Chern numbers of the two lowest bands $\nu_{1,2}$. We verified that this leads to a simultaneous and satisfactory measurement of these two topological numbers: For the data shown in Fig.3b of the main text we obtain $\nu_1=1.21(14)$, $\nu_2=-2.7(5)$ (black data points) and $\nu_1=1.04(10)$, $\nu_2=-2.2(3)$ (gray data points), where we have used the measured band populations $\eta_{\mu}(t)$ in Fig.~3c of the main text. \\

In Figure~\ref{fig:berry_curvature} we show the energy spectrum and the Berry curvature distribution of the lowest band for the homogeneous Harper-Hofstadter Hamiltonian ($\delta=0$) in eq.~(\ref{ideal_harper_stag}). In the experiment, the additional staggered detuning $\delta$ triggers a topological phase transition at $\delta=2J$ (Fig.4b in the main text and Sect.~\ref{sect:theorygauge}). For $\delta>2J$, the bands are all topologically trivial, with Chern numbers $\nu_{\mu}=0$, as depicted in Fig.~\ref{fig:berry_curvature}b for $\delta=4J$. 

\subsection*{On the particle-hole symmetry and the Chern numbers}

The discrete Schr\"odinger equation associated with the Harper-Hofstadter model can be written in the form~\cite{Hofstadter,Kohmoto}
\begin{align}
E & \psi (m,n) = -J \bigl[ \psi (m+1,n) + \psi (m-1,n) \notag \\
&+ e^{i 2 \pi \alpha m}\psi (m,n+1) + e^{-i 2 \pi \alpha m}\psi (m,n-1)     \bigr ],\label{harper_eq}
\end{align}
where we explicitly used the Landau gauge $\bs B = B (0,x)$, and we stress that the following discussion does not depend on this choice. For $\alpha=p/q$, the system has a discrete translational invariance associated with the translation operators
\be
\hat T_x^q \psi (m,n)= \psi (m+q,n) , \quad \hat T_y \psi (m,n)=\psi (m,n+1).\notag
\ee
Using Bloch's theorem, we write the wave function as
\be
 \psi (m,n)=e^{i k_x m} e^{i k_y n} u_{\bs k} (m) , \quad u_{\bs k}(m+q)=u_{\bs k}(m),\notag
\ee
where the periodic function $u_{\bs k}(m)$ satisfies the equation
\begin{align}
E u_{\bs k}(m) = -J \bigl [ &u_{\bs k}(m+1) e^{i k_x} + u_{\bs k}(m-1) e^{-i k_x} \notag \\
&+ 2 \cos \left (2 \pi \alpha m +k_y \right ) u_{\bs k}(m)     \bigr ]. \label{harper_periodic}
\end{align}
The energy spectrum splits into several subbands $E_{\mu}(\bs k)$, where $\mu$ is the band index. Let us focus on a specific band $E_{\mu}(\bs k)$, located around the value $E^* <0$. The Chern number associated with this band is given by eq.~\eqref{chern}, namely,
\begin{align}
&\nu_{\mu}=\frac{1}{2 \pi} \int_{\text{FBZ}} \Omega_{\mu} (\bs k) \text{d}^2 k.\label{chern_2}
\end{align}

Now, let us show that the Chern number associated with the ``top band" $E_{\tilde{\mu}}(\bs k)$, located around the value $(-E^*) >0$ is the same as the Chern number of the ``bottom  band" $E_{\mu}(\bs k)$, considered above: $\nu_{\mu}=\nu_{\tilde{\mu}}$. We start with the Harper equation \eqref{harper_eq} and consider the transformation
\be
\psi (m,n) \rightarrow \tilde{\psi} (m,n) = (-1)^{m+n} \psi (m,n) .\label{transf}
\ee
The new functions satisfy the Harper equation, 
\begin{align}
(-E) &\tilde \psi (m,n) = -J \bigl[ \tilde \psi (m+1,n) + \tilde \psi (m-1,n) \notag \\
&+ e^{i 2 \pi \alpha m} \tilde \psi (m,n+1) + e^{-i 2 \pi \alpha m} \tilde \psi (m,n-1)     \bigr ],\label{harper_eq_2}
\end{align}
which is the same as in eq.~\eqref{harper_eq}, but with $E \rightarrow -E$. This latter result illustrates the particle-hole symmetry in the system, i.e., the fact that if there exists a state at $E$, then there necessarily exists a state at the opposite energy $-E$ [which can be traced back to the fact that the square lattice is bipartite]. As before, we write the solution of eq.~\eqref{harper_eq_2} as
\be
 \tilde \psi (m,n)=e^{i k_x m} e^{i k_y n} \tilde u_{\bs k} (m) , \quad \tilde u_{\bs k}(m+q)=\tilde u_{\bs k}(m),\notag
\ee
where the periodic functions $\tilde u_{\bs k}(m)$ satisfy the equation
\begin{align}
E \tilde u_{\bs k}(m) = -J \bigl [ & \tilde u_{\bs k}(m+1) e^{i (k_x+\pi)} + \tilde u_{\bs k}(m-1) e^{-i (k_x+ \pi)} \notag \\
&+ 2 \cos \left (2 \pi \alpha m +(k_y+ \pi) \right ) \tilde u_{\bs k}(m)     \bigr ].
\end{align}
Comparing with eq.~\eqref{harper_periodic}, we find that the eigenstates associated with the ``top band" $E_{\tilde{\mu}}(\bs k)$ [located around $(-E^*) >0$] can be obtained from the eigenstates associated with the ``bottom band" $E_{\mu}(\bs k)$ [located around $E^* <0$] through the relation
\be
u_{k_x,k_y}(m) = \tilde u_{k_x + \pi , k_y + \pi}(m).
\ee
In other words, the ``particle-hole" transformation in eq.~\eqref{transf}, which transforms a state of energy $E$ into a state of opposite energy $(-E)$, is also associated with the transformation $\bs k \rightarrow (k_x + \pi, k_y + \pi)$. Consequently, the Berry's curvatures associated with the two opposite bands are related by
\be
\Omega_{\mu} (k_x, k_y) = \Omega_{\tilde{\mu}} (k_x + \pi, k_y + \pi), \label{shift}
\ee
namely, both bands share the same curvature, up to an overall shift in the Brillouin zone. 


As a corollary, two opposite bands of the Harper-Hofstadter model [located around $E^*$ and $(-E^*)$, respectively] necessarily share the same Chern number [see eq.~\eqref{chern}], as this quantity averages the Berry's curvature over the FBZ [i.e. the overall shift in eq.~\eqref{shift} can be eliminated by a redefinition of the FBZ]. 

\section{Experimental sequence}
In this section we describe the experimental sequence used for the transverse Hall drift and band population measurements.
\paragraph*{\textbf{Loading sequence}} The experimental sequence started by loading a Bose-Einstein condensate of $^{87}$Rb atoms within $150\,$ms into a two-dimensional optical superlattice. Along each of the axes two standing waves were superimposed with $\lambda_s=767\,$nm and $\lambda_L=2\lambda_s$. The relative phase between them was chosen such that a lattice potential with staggered energy offsets $\Delta+\delta_x$ along $x$ and $\delta_y$ along $y$, with $\delta_x\approx \delta_y \equiv \delta$ and $\delta < \Delta$, was created. The lattice depths were $V_{y}=10(1)\,$Er$_s$, $V_{yL}=1.75(5)\,$Er$_L$, $V_{x}=6.0(2)\,$Er$_s$ and $V_{xL}=5.25(16)\,$Er$_L$, with Er$_{\alpha}=\hbar^2 k_{\alpha}^2/(2m),  \alpha=\{s,L\}$. At this point of the sequence all atoms were loaded into the low energy sites. The two pairs of beams used for the modulation were then switched on in $30\,$ms, with a frequency difference $\omega_r=-\omega_b=\pm\Delta / \hbar$; at this stage, no resonant tunneling between neighboring sites was induced, due to the offset detuning $\delta$. After that, we ramped down the long lattices within $30\,$ms to $V_{yL}=0\,$Er$_L$ and $V_{xL}=3.25(10)\,$Er$_L$, which corresponds to $\delta=0$. For these values, resonant laser-assisted tunneling along $x$ and bare tunneling along $y$ occurred, simultaneously creating a homogeneous flux $\Phi=\pm \pi/2$ (aligned along $-\hat{\mathbf{e}}_z$) depending on the sign of the modulation frequency. We checked that all lattice sites were equally populated after the loading sequence.\paragraph*{\textbf{Loading sequence for lattice setup with trivial topology}}
The loading sequence described above is also used to load the atoms into the staggered flux lattice. The only difference is that the modulation is switched on with a frequency difference $\omega_r=\omega_b=\pm\Delta / \hbar$, which results in a flux $\Phi=\pm(-1)^m\pi/2$. \\
The sequence for the lattice without flux was performed in a similar manner. It started by loading the atoms into a two-dimensional superlattice with lattice depths $V_{x}=V_{y}=10(1)\,$Er$_s$, $V_{xL}=V_{yL}=1.75(5)\,$Er$_L$ within $150\,$ms. Subsequently the long lattices were decreased to zero within $30\,$ms.  In this way a simple square lattice configuration without flux and a tunnel coupling of $J/h=75(3)\,$Hz along both directions was realized.
\paragraph*{\textbf{Optical gradient}}
The optical gradient used to induce Bloch oscillations along $y$ was realized using an additional laser beam with wavelength $\lambda_z=844\,$nm. It was focused at the atom position to a waist of about $125\,\mu$m and aligned such that the atomic cloud is located at the maximum slope of the Gaussian beam profile along $y$. Along $x$ the beam was centered on the atom position. The strength of the gradient was determined through independent measurements of Bloch oscillations in a one-dimensional lattice with $V_{y}=10(1)\,$Er$_s$.
\paragraph*{\textbf{Band mapping sequence}}
To measure the populations in different Hofstadter bands, we reversed our loading sequence described above and ramped up the long lattices to $V_{yL}=1.75(5)\,$Er$_L$ and $V_{xL}=5.25(16)\,$Er$_L$, respectively, within $30\,$ms. At this point of the sequence, tunneling is off-resonant along both directions, and the Hofstadter bands map onto the bands of the usual 2D superlattice. We then suddenly switched off the modulation and used standard detection techniques to infer the momentum distribution and band populations.

\begin{figure}[t!]
\begin{center}
\includegraphics{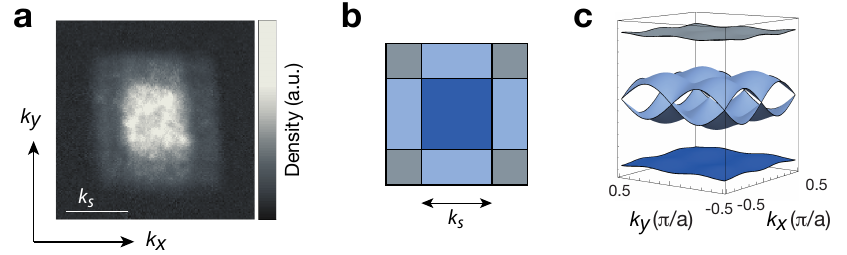}
\caption{Momentum distribution of the atoms in the three well-separated Hofstadter bands. \textbf{a} This data was obtained after applying the loading and band-mapping sequence as described in the Method section of the main text. Here we show an average of 15 independent measurements obtained after $10\,$ms time-of-flight. \textbf{b} Schematic illustration of the corresponding Brillouin zones. \textbf{c} Energy spectrum of the Harper-Hofstadter model for $\Phi=\pi/2$. The color code illustrates the connection between the Brillouin zones and the Hofstadter bands.}\label{fig:S5}
\end{center}
\end{figure}

\section{Momentum distribution in the Hofstadter bands}
\label{sect:bandmapping}

\begin{figure}[t!]
\begin{center}
\includegraphics{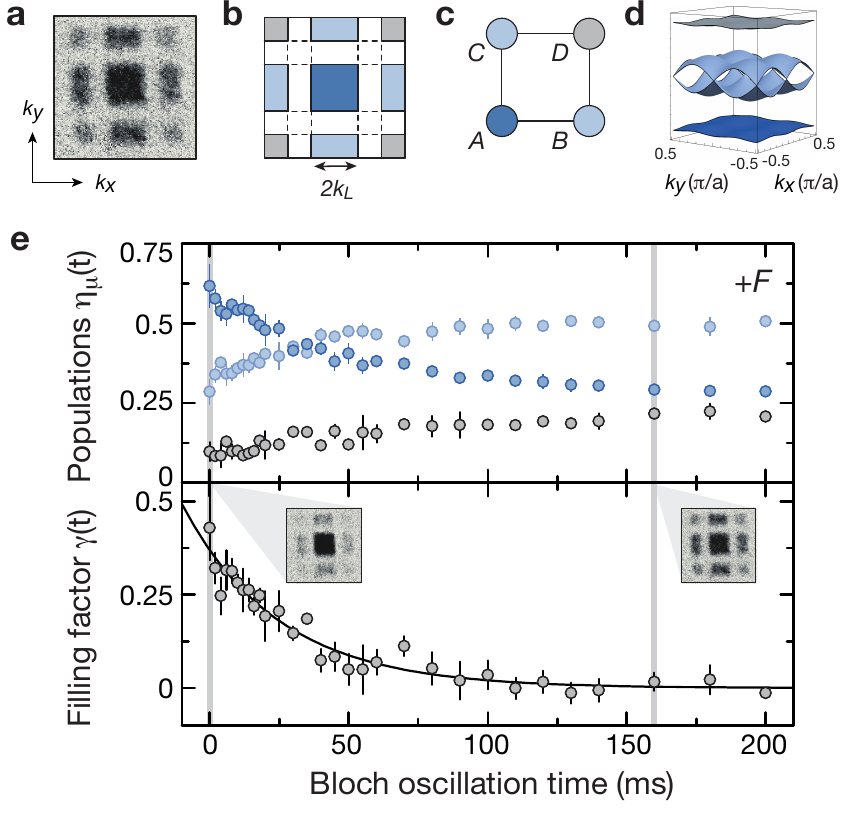}
\caption{Determination of the populations $\eta_{\mu}(t)$ and the corresponding filling factor $\gamma(t)$ in the three different Hofstadter bands. 
\textbf{a} Typical experimental image obtained after mapping the populations of the Hofstadter bands onto higher Bloch bands of the static two-dimensional superlattice as described in Sect.~\ref{sect:populations}. The momentum distribution was measured after  $10\,$ms time-of-flight. \textbf{b} Schematic drawing of the corresponding Brillouin zones. \textbf{c} Illustration of the four non-equivalent sites of the $2a\times 2a$ unit cell, whose dimensions are preserved during the sequence. \textbf{d} Energy spectrum of the Harper-Hofstadter model for $\Phi=\pi/2$. \textbf{e} Evolution of the band populations $\eta_{\mu}(t)$  as displayed in Fig.~3c of the main text and the corresponding filling factor $\gamma(t)$ for $Fa/h=38.4(8) \mathrm{Hz}$. The solid line in the lower panel shows an exponential fit to our data, which was used to extract the Chern number $\nu_{\text{exp}}$. The color code illustrates the connection between the Brillouin zones, sites in the unit cell, the Hofstadter bands and the corresponding measured band populations $\eta_{\mu}(t)$. The insets in the lower panel show typical experimental images obtained after the band-mapping sequence.}\label{fig:populations}
\end{center}
\end{figure}

The Hall drift as a response to an externally applied force given in eq.~(3) of the main text is based on the assumption that the distribution of bosonic atoms in the different Hofstadter bands is incoherent and homogeneous. To check this assumption independently in the experiment we mapped the band populations of the Hofstadter bands onto the ones of a static two-dimensional superlattice. This is achieved by reversing the loading sequence illustrated in Fig.~2 of the main text (see also Methods). After loading the atoms into the three well-separated Hofstadter bands, a staggered detuning $\delta$ is ramped up within $30\,$ms along both directions to suppress tunneling in the presence of the modulation (see main text and Methods). Switching off the modulation then maps the populations in the bands of the explicitly time-dependent Hamiltonian onto the ones of the static superlattice potential. The size of the Brillouin zone and thus the number of bands remains unchanged during the mapping sequence, such that the population of different $k$-states is preserved if scattering processes and heating effects during the ramp are neglected. Consequently the momentum distribution in the bands of the static superlattice potential reflects the one of the Hofstadter bands. All fields are then switched off adiabatically to map the quasimomentum distribution onto the real-space momentum distribution. After letting the atoms expand for $10\,$ms we measured the distribution using standard absorption imaging. The result is shown in Fig.~\ref{fig:S5}a. The connection to the Brillouin zones and the bands of the Harper-Hofstadter Hamiltonian are illustrated by the schematic drawings in Fig.~\ref{fig:S5}b,c. It can be seen that the atoms are distributed homogeneously over the lowest band. A fraction of atoms also populates the higher bands in a homogeneous manner. Our data is thus consistent with our assumption of homogeneous populations in the bands.

\section{Band-population measurement}
\label{sect:populations}

The population in different Hofstadter bands can be measured using the sequence described in the previous section by counting the atoms in the different Brillouin zones. The zones are, however, connected which complicates the precise evaluation of the corresponding atom numbers. To simplify the counting we apply a slightly different sequence which allows us to transfer the atoms to higher Bloch bands such that they appear in well separated Brillouin zones. We first map the band populations of the Hofstadter bands onto the ones of the static two-dimensional lattice without periodic driving (see previous section). After having switched off the periodic modulation the staggered energy offset between neighboring sites is given by $\delta + \Delta$ along $x$ and by $\delta$ along $y$. If these energy offsets are large enough compared to the bare couplings $J_x$ and $J_y$, tunneling is suppressed and the populations in different Bloch bands correspond to populations on different sites $N_q$, $q=A,B,C,D$ (see Fig.~\ref{fig:populations}). These can be measured by transferring the populations on different sites to higher Bloch bands and performing a subsequent band-mapping technique~\cite{Nascimbene}. The connection between the Hofstadter bands, sites in the unit cell and Brillouin zones is illustrated by the color code in Fig.~\ref{fig:populations}. Using the measured site-populations $N_q$ we evaluated the occupation in the three Hofstadter bands, $\eta_1=N_A/N_{\text{tot}}$, $\eta_2=(N_B+N_C)/N_{\text{tot}}$ and $\eta_3=N_D/N_{\text{tot}}$, where $N_{\text{tot}}$ is the total atom number (Fig.~\ref{fig:populations}e). By fitting an exponential to the corresponding filling factor $\gamma(t)=\eta_1(t)-\eta_2(t)+\eta_3(t)$ we determined the Chern number $\nu_{\text{exp}}$ from the measured differential shift $2x(t)$ according to equation~(5) in the main text, where $\nu_1$ was the only free fit parameter.

\section{Absolute center-of-mass positions}

\begin{figure}[t!]
\begin{center}
\includegraphics{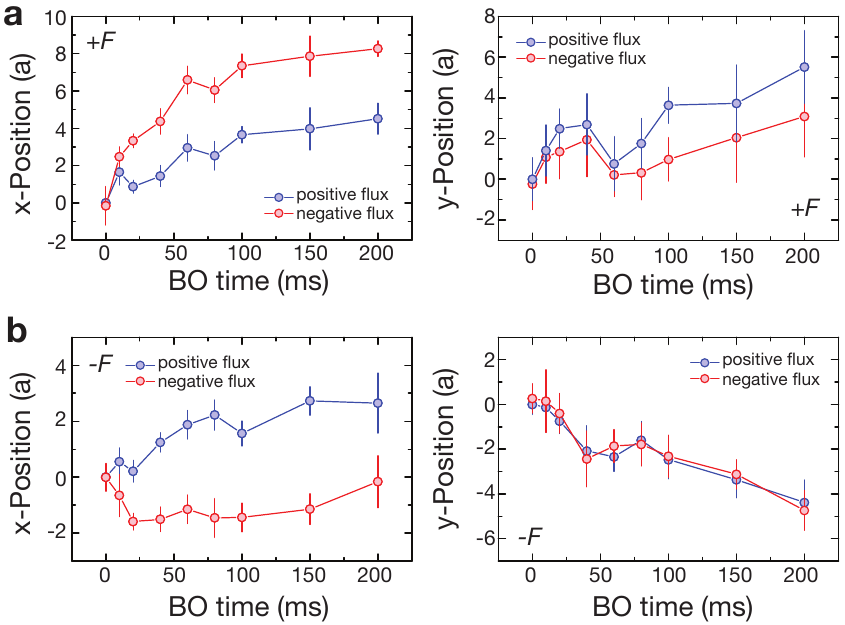}
\caption{Absolute center-of-mass positions along $x$ and $y$ as a function of the Bloch oscillation (BO) time for a gradient $\mathbf{F}=F \hat{\mathbf{e}}_y$ \textbf{(a)} and $\mathbf{F}=-F \hat{\mathbf{e}}_y$ \textbf{(b)} corresponding to the data shown in Fig.~3 of the main text.}\label{fig:S6}
\end{center}
\end{figure}

To study the Hall deflection of the cloud we evaluated the differential shift for positive and negative flux $\Phi$ by subtracting the center-of-mass (COM) positions of the cloud as depicted in Fig.~3a,b of the main text. In this section we show the absolute COM positions of the cloud for positive (blue) and negative (red) sign of the flux corresponding to the black (Fig.~\ref{fig:S6}a) and gray data points (Fig.~\ref{fig:S6}b) in Fig.~3b of the main text. We observe a clear splitting of the positions perpendicular ($x$-direction) to the gradient for both directions $\pm F \hat{\mathbf{e}}_y$. The additional drift in the $x$-position and differential shift in the $y$-position for $\mathbf{F}=F \hat{\mathbf{e}}_y$ (Fig.~\ref{fig:S6}a) is most likely due to a slight misalignment of the gradient such that it has a small component along $x$. We checked that this can be removed by aligning the optical gradient (see Method section of the main text) more carefully as can be seen in  Fig.~\ref{fig:S6}b, where the data was taken just after aligning the optical gradient. The global drift along $x$ in this case is absent as well as the differential shift along $y$. Additionally we checked that the differential shift perpendicular to the gradient is not affected.

\section{Numerical simulations and short-time dynamics}

In this Section, we discuss the methods used to simulate the dynamics. In the regime where the band populations $\eta_n(t)=\eta_n^0$ are constant, which is expected for sufficiently weak forces and short-time dynamics, the center-of-mass displacement follows the equations of motion,
\begin{align}
 x(t)  = x(t_0) - \frac{2 F t  a^2}{\pi \hbar} \gamma_0 \nu_1, \quad \gamma_0=   \eta_1^0  - \eta_2^0+ \eta_3^0 , \label{com-dis_2_bis}
\end{align}
as already discussed in Section~\ref{section:anomalous}. For a given force $F$ and initial band fillings $\eta_n^0$, this simple equation describes the ``ideal" linear drift of the cloud. 

In order to gain more insight on the dynamics captured by the effective Harper-Hofstadter Hamiltonian, and hence to verify the  validity of eq.~\eqref{com-dis_2_bis}, we have simulated the full non-interacting problem. In the absence of the force $F$, we write the Hamiltonian ruling the dynamics as [eq.~\eqref{sim_ham_new}]
\begin{align}
\hat H_{\rm eff}  &=   -J \sum_{m,n} \bigl \{ \hat a_{m+1,n}^{\dagger} \hat a_{m,n} e^{i \left [ \pi/2 (m+n) - \phi_0   \right ]}  +  \text{h.c.}  \notag \\
& \qquad \qquad + (1+f_{m,n}) \hat a_{m,n+1}^{\dagger} \hat a_{m,n} + \text{h.c.}\bigr \} \notag\\
&+\frac{\delta}{2} \sum_{m,n} \left [(-1)^{m} \!+\! (-1)^{n} \right ] \hat n_{m,n} + \hat V_{\text{conf}} \label{sim_ham} ,
\end{align}
which corresponds to the first-order effective Hamiltonian in eq.~\eqref{final_ham} together with the main second-order corrections $f_{m,n}$ defined in eq.~\eqref{inhom_y_corr_bis}. This Hamiltonian corresponds to the Harper-Hofstadter Hamiltonian with flux $\Phi=\pi/2$, and it includes the staggered potential detuning, and the external harmonic potential $\hat V_{\text{conf}}$, see Section~\ref{sect:effectiveHam}. In the following, we take $\hat V_{\text{conf}}=\sum_{m,n} V(m,n) \hat n_{m,n}$, with the experimental configuration $V(m,n)= \beta (0.5 m^2 + n^2)$ and $\beta=10^{-3}J$. To simulate the dynamics in the presence of a force, we follow the strategy described in Ref.~\cite{Dauphin}. We first establish the initial condition by confining the system within a certain radius $r_0\sim 10-30a$, using a  potential $\hat V_{\text{initial}}$; to simplify the analysis, we take an abrupt circular potential $\hat V_{\text{initial}}\sim (r/r_0)^{\zeta}$ with $\zeta \gg 10$, but we note that smoother potentials could also be considered for the initial preparation~\cite{Dauphin}. We diagonalize the corresponding Hamiltonian matrix  $\hat H_{\text{initial}}=\hat H_{\rm eff} + \hat V_{\text{initial}}$  on a finite system of radius $r > r_0$, and we classify its eigenstates $\chi_{\alpha}$ in terms of the three bulk bands, based on their energies $E_{\alpha}$; we note that the three-band structure clearly appears in the density of states. Having established the initial population of the bands, we then compute the time evolution of each state $\chi_{\alpha} (t)$ according to the Hamiltonian $\hat H_{\text{evol}}=\hat H_{\rm eff} + \hat V_{\text{force}}$, where $\hat V_{\text{force}}= - F \hat y$ describes the force acting on the particles at time $t>0$ along the $y$ direction; on the lattice, the operator $\hat y$ is defined as $\hat y = a \sum_{m,n} n \, \hat n_{m,n}$. The center-of-mass displacement $x(t)$ is then evaluated by computing the spatial density $\rho(\bs x,t)$, which is obtained from the populated evolving states $\chi_{\alpha} (t)$. \\

\begin{figure}[h]
\begin{center}
\includegraphics{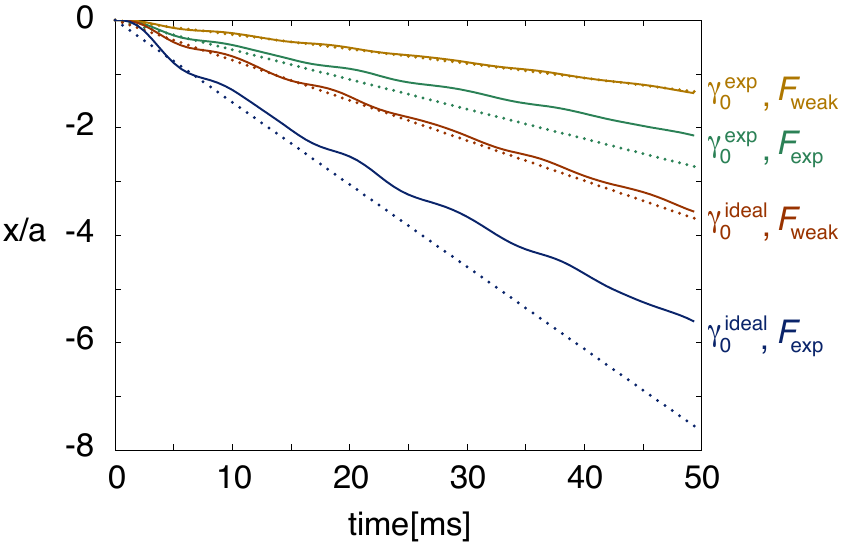}
\caption{Numerical simulations showing the center-of-mass displacement $x(t)$, in the direction perpendicular to the force $\bs F = F \hat{\bs{e}}_y$. The dynamics is governed by the Harper-Hofstadter Hamiltonian in eq.~\eqref{sim_ham}. The full curves correspond to different initial band populations and force strengths: the perfect filling of the lowest band is $\gamma_0^{\text{ideal}}=1$, the experimental filling is $\gamma_0^{\text{exp}}=0.36$, the ``weak" force is $F_{\text{weak}}=0.25 J/a$ and the experimental force is $F_{\text{exp}}=0.52 J/a$. The dotted lines show the linear drift predicted by eq.~\eqref{com-dis_2_bis} for each of the four situations; these linear trajectories are valid in the regimes where the band populations are constant (i.e. weak forces or short times). In all cases, we set $\delta=0$, $\phi_0=\pi/4$ and $\kappa/(\hbar \omega) = 0.58$, so that the effective tunneling along $y$ is $J_y^{\text{eff}} \approx 0.83 J$, and $J_x^{\text{eff}}=J$. As shown by the green shaded area in Fig.~3a of the main text, the trajectories are found to be similar for the wide range $\phi_0 \in [0 , \pi]$. }\label{fig:S8bis}
\end{center}
\end{figure}

Numerical results are illustrated in Fig.~\ref{fig:S8bis}, where the center-of-mass displacement $x(t)$ is plotted for different initial band populations $\gamma_0$ and force strengths $F$. First, let us consider a situation where the force is reasonably weak, $F_{\text{weak}}=0.25 J/a$. In this case, the motion is found to follow the linear behavior predicted by \eqref{com-dis_2_bis}, and the band populations remain approximatively constant for long times, $\gamma (t)=\gamma_0$. When the atoms populate the lowest band only, $\gamma_0^{\text{ideal}}=1$, they all undergo a net drift along the $-x$ direction due to the positive Berry curvature associated with this band (see eq.~\eqref{ano_vel} and Fig.~\ref{fig:berry_curvature}). For a typical experimental filling $\gamma_0^{\text{exp}}=0.36$, about $30 \%$ of the atoms populate the central band (with Chern number $\nu_2=-2$), and they propagate in the opposite direction $+x$: this leads to a slower center-of-mass velocity, captured by the factor $\gamma_0<1$ in the equations of motion \eqref{com-dis_2_bis}. Then, when the force is stronger, $F_{\text{exp}}=0.52 J/a$, as it is the case in the experiment to optimize the displacement measurement, $\gamma (t)$ is no longer constant and the simulations show a clear deviation from the linear motion \eqref{com-dis_2_bis} for times $t \sim 20\,\text{ms}$ (both for $\gamma_0=1$ and $\gamma_0<1$). A comparison with our experimental data for times $t\leq 35\,$ms is shown in Fig.~3a (main text), indicating a good agreement in the short-time regime. This short-time analysis, together with the measured initial band populations $\eta^0_{\mu}=\{0.55(6),0.31(3),0.13(3)\}$, and the equations of motion in eq.~\eqref{com-dis_2_bis}, provides a reasonable experimental value for the Chern number of the lowest band $\nu_{\exp}=0.9(2)$. The green shaded area in Fig.~3a (main text) delimits the numerically simulated trajectories obtained in the range $\phi_0 \in [0 , \pi]$, using the measured initial band populations ($\gamma_0 \approx 0.36$), and other experimental parameters ($\delta=0$, $\kappa/(\hbar \omega) = 0.58$). As already illustrated in Fig.~\ref{fig:S8bis}, the simulations already show a deviation from the linear behavior for times $t \sim 20\,\text{ms}$; this explains the reduced value of the experimentally determined Chern number, which includes data points up to $t=35\,$ms. We point out that the linear-motion breakdown (i.e. the Landau-Zener induced inter-band transitions) signaled by our simulations only  constitutes a partial explanation for the Hall drift saturation observed in the experiments. Indeed, we expect that additional effects, which are not captured by the present simulations (such as interactions and heating processes), could potentially lead to stronger band repopulation.

\end{document}